\documentclass[aps,prb,twocolumn,showpacs]{revtex4}
\usepackage{epsfig,graphicx}
\usepackage{amssymb}
\begin{document}
\title{Implementation of screened hybrid functionals based on the Yukawa
potential within the LAPW basis set}
\author{Fabien Tran}
\author{Peter Blaha}
\affiliation{Institute of Materials Chemistry, Vienna University of Technology,
Getreidemarkt 9/165-TC, A-1060 Vienna, Austria}

\begin{abstract}

The implementation of screened hybrid functionals into
the \textsc{wien2k} code, which is based on the LAPW basis set, is reported.
The Hartree-Fock exchange energy and potential are screened by means
of the Yukawa potential as proposed by Bylander and Kleinman
[Phys. Rev. B \textbf{41}, 7868 (1990)] for the calculation of
the electronic structure of solids with the screened-exchange
local density approximation. Details of the formalism, which is based on the
method of Massidda, Posternak, and Baldereschi
[Phys. Rev. B \textbf{48}, 5058 (1993)] for the unscreened
Hartree-Fock exchange are given.
The results for the transition-energy and structural properties
of several test cases are presented.
Results of calculations of the Cu electric-field gradient in
Cu$_{2}$O are also presented, and it is shown that the hybrid functionals
are much more accurate than the standard local-density or generalized
gradient approximations.

\end{abstract}

\maketitle

\section{\label{Introduction}Introduction}

Until now, most Kohn-Sham (KS) density functional theory (DFT)
\cite{HohenbergPR64,KohnPR65} calculations on solids
have been done using either the local density approximation (LDA) or
the generalized gradient approximation (GGA) for the exchange-correlation
energy. Calculations were done exclusively with LDA\cite{KohnPR65} until the
early 90s, when the GGA functional PW91\cite{PerdewPRB92b} was proposed and
then implemented in computer codes for solid-state calculations.
A few years later, a GGA functional with a simpler analytical form than PW91,
namely PBE,\cite{PerdewPRL96} but giving nearly
identical results has been proposed and is nowadays the standard functional.
The successes of the semilocal LDA and GGA
approximations rely on the fact that the accuracy is usually good enough to be
useful, in particular for the calculation of the geometrical parameters
and other quantities like the bulk modulus or the phonon spectrum.
(See, e.g., Refs. \onlinecite{TranPRB07} and \onlinecite{HaasPRB09} for
a compilation of lattice constants and bulk moduli calculated with various
GGA functionals.) However, it is known that there are classes of systems,
e.g., strongly correlated or van der Waals systems, whose properties are
not described properly by semilocal functionals already at the qualitative
level.

It is also well known that the KS band gap, defined as the conduction
band minimum (CBM) minus the valence band maximum (VBM),
obtained from a semilocal functional is much smaller than the experimental
band gap (defined as the ionization potential $I$ minus the
electron affinity $A$). However, it is important to note that
this problem, known as the \textit{band gap problem}, is more general and has its
roots in the KS-DFT method itself, and actually the KS band gap
calculated with the \textit{exact multiplicative} KS potential would
differ from $I-A$ by the derivative discontinuity $\Delta_{\text{xc}}$ of the
exchange-correlation potential (see Ref. \onlinecite{KummelRMP08} for a review).
Since $\Delta_{\text{xc}}$ can be of the same order as the KS band gap, the
\textit{exact} KS band gap can differ substantially from $I-A$.

There are several methods to obtain orbital energies which lead to values for
$\text{CBM}-\text{VBM}$ comparable to $I-A$.
If one wants to stay inside the true KS framework (i.e., KS equations with
a multiplicative potential), exact exchange (EXX) calculations
(see, e.g., Refs. \onlinecite{StadelePRB99} and \onlinecite{EngelPRL09})
or advanced semilocal potentials\cite{TranPRL09} can do a good job.
Alternatively one can
use a non-multiplicative potential, which means to use a method which
lies outside the KS framework, but belongs to the so-called generalized KS
framework.\cite{SeidlPRB96} Most of these methods
mix the DFT and Hartree-Fock (HF) theories and the best known are the
LDA+$U$,\cite{AnisimovPRB91} screened-exchange LDA (sX-LDA),
\cite{BylanderPRB90} and hybrid\cite{BeckeJCP93a,BeckeJCP93b} methods.
The $GW$ method can yield very accurate band structures,
in particular if it is applied self-consistently,
but it is a very expensive method (see Ref. \onlinecite{BechstedtPSSB09} for
a review).

The LDA+$U$ method (see Ref. \onlinecite{YlvisakerPRB09} for a recent review)
consists of applying an approximate (but very cheap) form of HF only to the
electrons which are not well described by semilocal functionals.
Typical examples are
the $3d$ or $4f$ electrons in strongly correlated systems
(e.g., transition-metal and rare-earth oxides) that are very localized and
hence lead to large self-interaction error when a semilocal functional is used
(this results in too small band gaps and magnetic moments).
In the sX-LDA method, the short-range (SR) part of LDA exchange is
replaced by the SR part of HF exchange, where the SR
part is defined by replacing the bare Coulomb potential by the screened Yukawa
potential\cite{YukawaPPMSJ35}
into the corresponding expressions for the energy and potential.
The sX-LDA method has been implemented within the pseudopotential plane-wave
\cite{BylanderPRB90,SeidlPRB96,ClarkPRB10,LeePRB07} and
linearized-augmented plane-wave \cite{AsahiPRB99,AsahiPRB00,GellerAPL01}
basis sets, and it has been shown that sX-LDA improves substantially over LDA
for the band gap of semiconductors and insulators.

Despite the fact that reports about the implementation of the HF method
in solid-state codes started to appear already in the 70s (see, e.g.,
Refs. \onlinecite{DagensPRB72,SvanePRB87,PisaniLNC88,MassiddaPRB93}),
it is only in the early 2000s that
the first calculations on solids with hybrid methods
were reported,\cite{BredowPRB00,MuscatCPL01,PerryPRB01} which
is much later than for molecules.\cite{BeckeJCP93a,BeckeJCP93b}
In hybrid methods, a certain percentage (between 10\% and 50\%) of semilocal
exchange is replaced by HF exchange, while the correlation remains purely semilocal.
As for molecules, the hybrid functionals have shown to lead to (much) better
results than semilocal functionals for various types of materials and properties.
In particular, they lead to band structures which are usually in good agreement
with experiment as shown for classical semiconductors and insulators
(see, e.g., Refs. \onlinecite{BredowPRB00,MuscatCPL01,CoraSB04}) and
strongly correlated materials (see, e.g.,
Refs. \onlinecite{BredowPRB00,PerryPRB01,MoreiraPRB02,KudinPRL02,CoraSB04,FranchiniPRB05}).
The most common hybrid functionals are B3LYP \cite{BeckeJCP93b,StephensJPC94}
and PBE0 \cite{ErnzerhofJCP99,AdamoJCP99} which contain 20\% and 25\% of HF
exchange, respectively. However, for solids the long-range (LR) nature of HF
exchange leads to technical difficulties.
For calculations done in real space, the results converge very slowly
with respect to the number of
neighboring unit cells that are taken into account for the calculation of HF
exchange, while for calculations done in reciprocal space, the slow convergence
is with respect to the number of $\mathbf{k}$-points for the integrations in
the Brillouin zone.

To reduce this problem of slow convergence, Heyd \textit{et al}. (HSE)
\cite{HeydJCP03,HeydJCP04a,HeydJCP04b}
proposed to consider only the SR part
of HF exchange (as done in sX-LDA), and therefore to keep 100\% of semilocal
LR exchange. This was done
by splitting the Coulomb operator into SR and LR components by using the
error function.\cite{AdamsonCPL96,Savin}
Since then, it has been shown that the HSE functional, which is
based on PBE0, leads to very good results for semiconductors and insulators
\cite{HeydJCP05,PaierJCP06,KrukauJCP06,WuPRB09,MarquesPRB11} including strongly
correlated systems, \cite{MarsmanJPCM08}
and several recent papers reporting the implementation of HSE have appeared.
\cite{PaierJCP06,GuidonJCP08} We also mention the onsite version
of HF exchange proposed by Nov\'{a}k \textit{et al}., \cite{NovakPSSB06} which
leads to very cheap calculations, but can be applied only to localized electrons.
This method has been used in the context of hybrid calculations.
\cite{TranPRB06,JolletPRB09}

In the present work, we report the implementation of screened hybrid functionals
into the \textsc{wien2k} code,\cite{WIEN2k} which is based on the full-potential
linearized augmented plane-wave plus local orbitals method
(abbreviated as LAPW in the following)
\cite{AndersenPRB75,Singh,SjostedtSSC00,MadsenPRB01} to solve the KS equations.
As done for the sX-LDA functional,\cite{BylanderPRB90} the HF exchange is
screened by means of the Yukawa potential in order to eliminate the LR HF
exchange. The calculation of the screened HF exchange is based on the
pseudocharge method \cite{WeinertJMP81} as proposed by Massidda \textit{et al}.
for the unscreened HF exchange.\cite{MassiddaPRB93}
In the papers of Asahi \textit{et al}. \cite{AsahiPRB99,AsahiPRB00} it is
mentioned that this mehod was used for the
implementation of the sX-LDA functional, but only very few details are given.
At this point we also mention Refs.
\onlinecite{NikolaevIJQC02,FriedrichCPC09,BetzingerPRB10,BetzingerPRB11}, in
which alternative ways of implementing the HF or EXX methods within the LAPW
basis set are presented.

The paper is organized as follows: in Sec. \ref{Screened}, the details of the
formalism of the unscreened and screened HF exchange for the LAPW basis set
are given and in Sec. \ref{hybrid},
the implemented screened hybrid functionals are presented.
In Sec. \ref{results}, the results for a few test cases and Cu$_{2}$O
are presented, and in Sec. \ref{summary} the summary of the work is given.

\section{\label{Screened}Screened Hartree-Fock exchange}

In this section, the formulas of the screened HF energy for the
LAPW basis set are given. The formulas are also valid (and implemented) for
the APW plus local orbitals basis set.\cite{SjostedtSSC00,MadsenPRB01}
For completeness and to allow comparison, the formulas for the unscreened case
are also given. For the Hamiltonian, only the basic formulas are given.
The LAPW method will not be described here, but details can be found in
Refs. \onlinecite{Singh,SjostedtSSC00,MadsenPRB01}.

\subsection{\label{Energy}Energy}

The HF exchange energy per unit cell (of volume $\Omega$)
is given by (all following equations are in Hartree atomic units)
\begin{equation}
E_{\text{x}}^{\text{HF}} = E_{\text{x,vv}}^{\text{HF}} +
E_{\text{x,vc}}^{\text{HF}} + E_{\text{x,cc}}^{\text{HF}},
\label{ExHF}
\end{equation}
where
\begin{widetext}
\begin{equation}
E_{\text{x,vv}}^{\text{HF}} =
-\frac{1}{2}\sum_{\sigma}\sum_{n,\mathbf{k},n',\mathbf{k}'}
w_{n\mathbf{k}}^{\sigma}w_{n'\mathbf{k}'}^{\sigma}
\int\limits_{\Omega}\int\limits_{\text{crystal}}
\psi_{n\mathbf{k}}^{\sigma*}(\mathbf{r})
\psi_{n'\mathbf{k}'}^{\sigma}(\mathbf{r})
v\left(\left\vert\mathbf{r}-\mathbf{r}'\right\vert\right)
\psi_{n'\mathbf{k}'}^{\sigma*}(\mathbf{r}')
\psi_{n\mathbf{k}}^{\sigma}(\mathbf{r}')
d^{3}r'd^{3}r,
\label{ExvvHF}
\end{equation}
\begin{equation}
E_{\text{x,vc}}^{\text{HF}} =
-\sum_{\sigma}\sum_{\alpha}^{\text{cell}}\sum_{n_{\text{c}},\ell_{\text{c}},m_{\text{c}}}
\sum_{n,\mathbf{k}}
w_{n\mathbf{k}}^{\sigma}
\int\limits_{\text{S}_{\alpha}}\int\limits_{\text{S}_{\alpha}}
\psi_{n\mathbf{k}}^{\sigma*}(\mathbf{r})
\psi_{n_{\text{c}}\ell_{\text{c}}m_{\text{c}}}^{\alpha\sigma}(\mathbf{r})
v\left(\left\vert\mathbf{r}-\mathbf{r}'\right\vert\right)
\psi_{n_{\text{c}}\ell_{\text{c}}m_{\text{c}}}^{\alpha\sigma*}(\mathbf{r}')
\psi_{n\mathbf{k}}^{\sigma}(\mathbf{r}')
d^{3}r'd^{3}r,
\label{ExvcHF}
\end{equation}
\begin{equation}
E_{\text{x,cc}}^{\text{HF}} =
-\frac{1}{2}\sum_{\sigma}\sum_{\alpha}^{\text{cell}}
\sum_{n_{\text{c}},\ell_{\text{c}},m_{\text{c}}\atop
n_{\text{c}}',\ell_{\text{c}}',m_{\text{c}}'}
\int\limits_{\text{S}_{\alpha}}\int\limits_{\text{S}_{\alpha}}
\psi_{n_{\text{c}}'\ell_{\text{c}}'m_{\text{c}}'}^{\alpha\sigma*}(\mathbf{r})
\psi_{n_{\text{c}}\ell_{\text{c}}m_{\text{c}}}^{\alpha\sigma}(\mathbf{r})
v\left(\left\vert\mathbf{r}-\mathbf{r}'\right\vert\right)
\psi_{n_{\text{c}}\ell_{\text{c}}m_{\text{c}}}^{\alpha\sigma*}(\mathbf{r}')
\psi_{n_{\text{c}}'\ell_{\text{c}}'m_{\text{c}}'}^{\alpha\sigma}(\mathbf{r}')
d^{3}r'd^{3}r,
\label{ExccHF}
\end{equation}
\end{widetext}
are the valence-valence (vv), valence-core (vc), and core-core (cc) terms,
respectively. In Eqs. (\ref{ExvvHF}) and (\ref{ExvcHF}),
$w_{n\mathbf{k}}^{\sigma}$ is the product of the $\mathbf{k}$-point weight
and the occupation number and $\psi_{n\mathbf{k}}^{\sigma}$
is a spin-$\sigma$ valence orbital of band index $n$ and wave vector
$\mathbf{k}$, whose LAPW basis set expansion in the interstitial
(I) and atomic spheres ($\text{S}_{\alpha}$) is given by
($\mathbf{r}_{\alpha}=\mathbf{r}-\tau_{\alpha}$, where $\tau_{\alpha}$ is the
position of nucleus $\alpha$)
\begin{equation}
\psi_{n\mathbf{k}}^{\sigma}(\mathbf{r}) =
\sum\limits_{\mathbf{K}}
c_{n,\mathbf{k}+\mathbf{K}}^{\sigma}
\phi_{\mathbf{k}+\mathbf{K}}^{\sigma}(\mathbf{r}),
\label{psink}
\end{equation}
\begin{equation}
\phi_{\mathbf{k}+\mathbf{K}}^{\sigma}(\mathbf{r}) =
\left\{
\begin{array}{l@{\quad}l}
\frac{1}{\sqrt{\Omega}}
e^{\text{i}\left(\mathbf{k}+\mathbf{K}\right)\cdot\mathbf{r}}, &
\mathbf{r}\in\text{I} \\
\sum\limits_{\ell,m}\sum\limits_{f}
d_{f,\mathbf{k}+\mathbf{K}}^{\alpha\sigma\ell m}
u_{f\ell}^{\alpha\sigma}(r_{\alpha})
Y_{\ell m}(\hat\mathbf{r}_{\alpha}), & \mathbf{r}\in\text{S}_{\alpha}
\end{array}
\right.,
\label{phikk}
\end{equation}
where $c_{n,\mathbf{k}+\mathbf{K}}^{\sigma}$ are the variational coefficients.
In the interstitial, the basis functions $\phi_{\mathbf{k}+\mathbf{K}}^{\sigma}$
are represented by plane-waves, while inside the atomic spheres,
$\phi_{\mathbf{k}+\mathbf{K}}^{\sigma}$ are linear combinations of 
products of radial functions $u_{f\ell}^{\alpha\sigma}$ and spherical
harmonics $Y_{\ell m}$. The coefficients
$d_{f,\mathbf{k}+\mathbf{K}}^{\alpha\sigma\ell m}$ are determined such
that the $\phi_{\mathbf{k}+\mathbf{K}}^{\sigma}$'s are continuous across the sphere
boundaries. For $f=1$, 2, and 3,
$u_{f\ell}^{\alpha\sigma}$ represents a radial function evaluated at a linearization
energy, its energy derivative evaluated at this same energy, and a
radial function evaluated at another linearization energy
(e.g., semicore states), respectively.
In Eqs. (\ref{ExvcHF}) and (\ref{ExccHF}),
$\psi_{n_{\text{c}},\ell_{\text{c}},m_{\text{c}}}^{\alpha\sigma}$
is a core orbital which is confined
inside the atomic sphere $\text{S}_{\alpha}$ and where
$n_{\text{c}}$, $\ell_{\text{c}}$, and $m_{\text{c}}$ are
the principal, azimuthal, and magnetic quantum numbers, respectively:
\begin{equation}
\psi_{n_{\text{c}}\ell_{\text{c}}m_{\text{c}}}^{\alpha\sigma}(\mathbf{r}) =
u_{n_{\text{c}}\ell_{\text{c}}}^{\alpha\sigma}(r_{\alpha})
Y_{\ell_{\text{c}}m_{\text{c}}}(\hat\mathbf{r}_{\alpha}).
\label{psinlm}
\end{equation}
In Eqs. (\ref{ExvvHF})-(\ref{ExccHF}), $v$ is either the unscreened potential
\begin{equation}
\frac{1}
{\left\vert\mathbf{r}-\mathbf{r}'\right\vert} =
\sum_{\ell=0}^{\infty}\sum_{m=-\ell}^{\ell}
\frac{4\pi}{2\ell+1}\frac{r_{<}^{\ell}}{r_{>}^{\ell+1}}
Y_{\ell m}^{*}\left(\hat{\mathbf{r}}\right)Y_{\ell m}\left(\hat{\mathbf{r}}'\right)
\label{vc}
\end{equation}
or the Yukawa screened potential \cite{YukawaPPMSJ35}
\begin{eqnarray}
\frac{e^{-\lambda\left\vert\mathbf{r}-\mathbf{r}'\right\vert}}
{\left\vert\mathbf{r}-\mathbf{r}'\right\vert} & = &
4\pi\lambda
\sum_{\ell=0}^{\infty}\sum_{m=-\ell}^{\ell}
i_{\ell}\left(\lambda r_{<}\right)k_{\ell}\left(\lambda r_{>}\right) \nonumber \\
& & \times
Y_{\ell m}^{*}\left(\hat{\mathbf{r}}\right)Y_{\ell m}\left(\hat{\mathbf{r}}'\right),
\label{vcs}
\end{eqnarray}
where $\lambda$ is the screening parameter and $i_{\ell}$ and $k_{\ell}$ are
spherical modified Bessel functions.\cite{Arfken}
Note that the spherical harmonics expansion of the screened potential
\cite{Arfken} is simpler than in
the case of the error function.\cite{AngyanJPA06}

\subsubsection{\label{VVT}Valence-valence term}

Following the idea of Massidda \textit{et al}. \cite{MassiddaPRB93}
the valence-valence term [Eq. (\ref{ExvvHF})] is cast into the following form
\begin{eqnarray}
E_{\text{x,vv}}^{\text{HF}} & = &
-\frac{1}{2}\sum_{\sigma}\sum_{n,\mathbf{k},n',\mathbf{k}'}
w_{n\mathbf{k}}^{\sigma}w_{n'\mathbf{k}'}^{\sigma} \nonumber \\
& & \times
\int\limits_{\Omega}\rho_{n\mathbf{k}n'\mathbf{k}'}^{\sigma}(\mathbf{r})
v_{n\mathbf{k}n'\mathbf{k}'}^{\sigma*}(\mathbf{r})d^{3}r,
\label{ExvvHF2}
\end{eqnarray}
where
\begin{equation}
\rho_{n\mathbf{k}n'\mathbf{k}'}^{\sigma}(\mathbf{r}) =
\psi_{n\mathbf{k}}^{\sigma*}(\mathbf{r})\psi_{n'\mathbf{k}'}^{\sigma}(\mathbf{r}),
\label{rhonk1}
\end{equation}
and
\begin{equation}
v_{n\mathbf{k}n'\mathbf{k}'}^{\sigma}(\mathbf{r}) =
\int\limits_{\text{crystal}}\rho_{n\mathbf{k}n'\mathbf{k}'}^{\sigma}(\mathbf{r}')
v\left(\left\vert\mathbf{r}-\mathbf{r}'\right\vert\right)d^{3}r'.
\label{vnk1}
\end{equation}
In the interstitial and spheres,
$\rho_{n\mathbf{k}n'\mathbf{k}'}^{\sigma}$ and
$v_{n\mathbf{k}n'\mathbf{k}'}^{\sigma}$ are expanded in Fourier and spherical
harmonics series, respectively (from now on, the index $\alpha$ of
the position $\mathbf{r}_{\alpha}$ from the nucleus $\alpha$ is suppressed
and we define $\mathbf{q}=\mathbf{k}'-\mathbf{k}+\mathbf{G}$):
\begin{equation}
\rho_{n\mathbf{k}n'\mathbf{k}'}^{\sigma}(\mathbf{r}) =
\left\{
\begin{array}{l@{\quad}l}
\sum\limits_{\mathbf{G}}\rho_{n\mathbf{k}n'\mathbf{k}'}^{\sigma\mathbf{G}}
e^{{\text{i}}\mathbf{q}\cdot\mathbf{r}},
& \mathbf{r} \in {\text{I}} \\
\sum\limits_{\ell,m}\rho_{n\mathbf{k}n'\mathbf{k}'}^{\alpha\sigma\ell m}(r)
Y_{\ell m}(\hat{\mathbf{r}}), & \mathbf{r} \in {\text{S}}_{\alpha} \\
\end{array}
\right.,
\label{rhonk2}
\end{equation}
\begin{equation}
v_{n\mathbf{k}n'\mathbf{k}'}^{\sigma}(\mathbf{r}) =
\left\{
\begin{array}{l@{\quad}l}
\sum\limits_{\mathbf{G}}
v_{n\mathbf{k}n'\mathbf{k}'}^{\sigma\mathbf{q}}
e^{{\text{i}}\mathbf{q}\cdot\mathbf{r}}, &
\mathbf{r} \in {\text{I}} \\
\sum\limits_{\ell,m}v_{n\mathbf{k}n'\mathbf{k}'}^{\alpha\sigma\ell m}(r)
Y_{\ell m}(\hat{\mathbf{r}}), & \mathbf{r} \in {\text{S}}_{\alpha} \\
\end{array}
\right..
\label{vnk2}
\end{equation}
In Eq. (\ref{rhonk2}),
$\rho_{n\mathbf{k}n'\mathbf{k}'}^{\sigma\mathbf{G}}$
are the Fourier coefficients of the periodic part of
$\rho_{n\mathbf{k}n'\mathbf{k}'}^{\sigma}$ and
$\rho_{n\mathbf{k}n'\mathbf{k}'}^{\alpha\sigma\ell m}$ is given by
\begin{equation}
\rho_{n\mathbf{k}n'\mathbf{k}'}^{\alpha\sigma\ell m}(r) =
\sum_{\ell_{1},\ell_{2}}\sum_{f_{1},f_{2}}
T_{\alpha\sigma n\mathbf{k}n'\mathbf{k}'}^{f_{1}f_{2}\ell_{1}\ell_{2}\ell m}
u_{f_{1}\ell_{1}}^{\alpha\sigma}(r)u_{f_{2}\ell_{2}}^{\alpha\sigma}(r),
\label{rhonklm}
\end{equation}
where
\begin{eqnarray}
T_{\alpha\sigma n\mathbf{k}n'\mathbf{k}'}^{f_{1}f_{2}\ell_{1}\ell_{2}\ell m} & = &
\sum_{m_{1}=-\ell_{1}}^{\ell_{1}}\sum_{m_{2}=-\ell_{2}}^{\ell_{2}}
C_{\ell_{1}m_{1}\ell m}^{\ell_{2}m_{2}} \nonumber \\
& & \times
\left(D_{\ell_{1}m_{1}}^{\alpha\sigma n\mathbf{k}f_{1}}\right)^{*}
D_{\ell_{2}m_{2}}^{\alpha\sigma n'\mathbf{k}'f_{2}}
\label{T}
\end{eqnarray}
with $C_{\ell_{1}m_{1}\ell m}^{\ell_{2}m_{2}}$ being Gaunt coefficients,
\begin{equation}
C_{\ell_{1}m_{1}\ell m}^{\ell_{2}m_{2}} =
\int\limits_{0}^{2\pi}\int\limits_{0}^{\pi}
Y_{\ell_{2}m_{2}}^{*}\left(\hat{\mathbf{r}}\right)
Y_{\ell_{1}m_{1}}\left(\hat{\mathbf{r}}\right)
Y_{\ell m}\left(\hat{\mathbf{r}}\right)
\sin\theta d\theta d\phi,
\label{Gaunt}
\end{equation}
and
$D_{\ell m}^{\alpha\sigma n\mathbf{k}f}=
\sum_{\mathbf{K}}c_{n,\mathbf{k}+\mathbf{K}}^{\sigma}
d_{f,\mathbf{k}+\mathbf{K}}^{\alpha\sigma\ell m}$.

$v_{n\mathbf{k}n'\mathbf{k}'}^{\sigma}$ is calculated by using
Weinert's method for solving the Poisson equation.\cite{WeinertJMP81}
(In Appendix \ref{basic}, a brief summary of Weinert's
method for the unscreened and screened potentials is given.)
For the unscreened and screened potentials, the Fourier coefficients
$v_{n\mathbf{k}n'\mathbf{k}'}^{\sigma\mathbf{q}}$
are given by
\begin{equation}
v_{n\mathbf{k}n'\mathbf{k}'}^{\sigma\mathbf{q}} =
4\pi\frac{\tilde{\rho}_{n\mathbf{k}n'\mathbf{k}'}^{\sigma\mathbf{q}}}
{\left\vert\mathbf{q}\right\vert^{2}}
\label{vnkG}
\end{equation}
and
\begin{equation}
v_{n\mathbf{k}n'\mathbf{k}'}^{\sigma\mathbf{q}} =
4\pi\frac{\tilde{\rho}_{n\mathbf{k}n'\mathbf{k}'}^{\sigma\mathbf{q}}}
{\left\vert\mathbf{q}\right\vert^{2} + \lambda^{2}},
\label{vnkGs}
\end{equation}
respectively, where
$\tilde{\rho}_{n\mathbf{k}n'\mathbf{k}'}^{\sigma\mathbf{q}}$
are the Fourier coefficients of the pseudocharge density
[see Eqs. (\ref{qqq})-(\ref{rho0s}) of Appendix \ref{formulasHF}].
Note that for the unscreened potential, the term corresponding to
$\mathbf{q}=\mathbf{0}$
(i.e., $\mathbf{k}=\mathbf{k}'$ and $\mathbf{G}=\mathbf{0}$) leads
to a singularity which has to be considered carefully (details are given
at the end of this section).

The radial function $v_{n\mathbf{k}n'\mathbf{k}'}^{\alpha\sigma\ell m}$ is
given by [$r_{<}=\min\left(r,r'\right)$, $r_{>}=\max\left(r,r'\right)$, and
$R_{\alpha}$ is the radius of the atomic sphere]
\begin{eqnarray}
v_{n\mathbf{k}n'\mathbf{k}'}^{\alpha\sigma\ell m}(r) & = &
\int\limits_{0}^{R_{\alpha}}
\rho_{n\mathbf{k}n'\mathbf{k}'}^{\alpha\sigma\ell m}(r')
G_{\ell}^{\alpha}\left(r,r'\right)r'^{2}dr' \nonumber \\
& & +
v_{n\mathbf{k}n'\mathbf{k}'}^{\alpha\sigma\ell m}(R_{\alpha})
P_{\ell}(r),
\label{vlm}
\end{eqnarray}
where $G_{\ell}^{\alpha}$ is Eq. (\ref{Gla}) and
$P_{\ell}=r^{\ell}/R_{\alpha}^{\ell}$ for the unscreened potential
or $G_{\ell}^{\alpha}$ is Eq. (\ref{Glas}) and
$P_{\ell}=i_{\ell}\left(\lambda r\right)/i_{\ell}\left(\lambda R_{\alpha}\right)$
for the screened potential.
In Eq. (\ref{vlm}),
\begin{equation}
v_{n\mathbf{k}n'\mathbf{k}'}^{\alpha\sigma\ell m}\left(R_{\alpha}\right) =
4\pi\text{i}^{\ell}
\sum_{\mathbf{G}}v_{n\mathbf{k}n'\mathbf{k}'}^{\sigma\mathbf{q}}
e^{{\text{i}}\mathbf{q}\cdot\tau_{\alpha}}
Y_{\ell m}^{*}\left(\widehat{\mathbf{q}}\right)
j_{\ell}\left(\left\vert\mathbf{q}\right\vert R_{\alpha}\right),
\label{vlmR}
\end{equation}
which is obtained by using the Rayleigh formula\cite{Arfken}
\begin{equation}
e^{\text{i}\mathbf{q}\cdot\mathbf{r}} =
4\pi\sum_{\ell=0}^{\infty}\sum_{m=-\ell}^{\ell}
\text{i}^{\ell}j_{\ell}\left(\left\vert\mathbf{q}\right\vert r\right)
Y_{\ell m}^{*}\left(\hat{\mathbf{q}}\right)Y_{\ell m}\left(\hat{\mathbf{r}}\right)
\label{eikr}
\end{equation}
in the Fourier expansion of
$v_{n\mathbf{k}n'\mathbf{k}'}^{\sigma}$ [Eq. (\ref{vnk2})], where
$j_{\ell}$ is a spherical Bessel function.\cite{Arfken}

$E_{\text{x,vv}}^{\text{HF}}$ is decomposed into its interstitial and atomic
sphere parts:
\begin{equation}
E_{\text{x,vv}}^{\text{HF}} = E_{\text{x,vv}}^{\text{HF,I}} +
\sum_{\alpha}^{\text{cell}}E_{\text{x,vv}}^{\text{HF},\text{S}_{\alpha}},
\label{ExvvHF3}
\end{equation}
where
\begin{eqnarray}
E_{\text{x,vv}}^{\text{HF,I}} & = &
-\frac{1}{2}\sum_{\sigma}\sum_{n,\mathbf{k},n',\mathbf{k}'}
w_{n\mathbf{k}}^{\sigma}w_{n'\mathbf{k}'}^{\sigma} \nonumber \\
& & \times
\int\limits_{\Omega}\rho_{n\mathbf{k}n'\mathbf{k}'}^{\sigma}(\mathbf{r})
v_{n\mathbf{k}n'\mathbf{k}'}^{\sigma*}(\mathbf{r})\Theta(\mathbf{r})d^{3}r \nonumber \\
& = &
-\frac{\Omega}{2}\sum_{\sigma}\sum_{n,\mathbf{k},n',\mathbf{k}'}
w_{n\mathbf{k}}^{\sigma}w_{n'\mathbf{k}'}^{\sigma} \nonumber \\
& & \times
\sum_{\mathbf{G}}
\left(\rho_{n\mathbf{k}n'\mathbf{k}'}^{\sigma}
v_{n\mathbf{k}n'\mathbf{k}'}^{\sigma*}\right)_{\mathbf{G}}
\Theta_{-\mathbf{G}},
\label{ExHFI}
\end{eqnarray}
with $\Theta(\mathbf{r})=1$ if $\mathbf{r}\in\text{I}$ and 0 if
$\mathbf{r}\in\text{S}_{\alpha}$, whose 
Fourier transform $\Theta_{\mathbf{G}}$ is given by
\begin{equation}
\Theta_{\mathbf{G}} =
\left\{
\begin{array}{l@{\quad}l}
-\frac{4\pi}{\Omega}\sum_{\alpha}^{\text{cell}}e^{-\text{i}\mathbf{G}\cdot\tau_{\alpha}}
R_{\alpha}^{3}\frac{j_{1}\left(\left\vert\mathbf{G}\right\vert R_{\alpha}\right)}
{\left\vert\mathbf{G}\right\vert R_{\alpha}}, &
\mathbf{G}\neq\mathbf{0} \\
1-\frac{4\pi}{3\Omega}\sum_{\alpha}^{\text{cell}}R_{\alpha}^{3}, & \mathbf{G}=\mathbf{0}
\end{array}
\right.,
\label{ThetaG}
\end{equation}
and
\begin{widetext}
\begin{eqnarray}
E_{\text{x,vv}}^{\text{HF},\text{S}_{\alpha}} & = &
-\frac{1}{2}\sum_{\sigma}\sum_{n,\mathbf{k},n',\mathbf{k}'}
w_{n\mathbf{k}}^{\sigma}w_{n'\mathbf{k}'}^{\sigma}
\int\limits_{\text{S}_{\alpha}}\rho_{n\mathbf{k}n'\mathbf{k}'}^{\sigma}(\mathbf{r})
v_{n\mathbf{k}n'\mathbf{k}'}^{\sigma*}(\mathbf{r})d^{3}r \nonumber \\
& = & -\frac{1}{2}\sum_{\sigma}\sum_{n,\mathbf{k},n',\mathbf{k}'}
w_{n\mathbf{k}}^{\sigma}w_{n'\mathbf{k}'}^{\sigma}\sum_{\ell,m}\left[
\sum_{\ell_{1},\ell_{2}\atop\ell_{3},\ell_{4}}\sum_{f_{1},f_{2}\atop f_{3},f_{4}}
T_{\alpha\sigma n\mathbf{k}n'\mathbf{k}'}^{f_{1}f_{2}\ell_{1}\ell_{2}\ell m}
\left(T_{\alpha\sigma n\mathbf{k}n'\mathbf{k}'}^{f_{3}f_{4}\ell_{3}\ell_{4}\ell m}
\right)^{*}\right. \nonumber \\
& &
\left.
\times
\int\limits_{0}^{R_{\alpha}}\int\limits_{0}^{R_{\alpha}}
u_{f_{1}\ell_{1}}^{\alpha\sigma}(r)u_{f_{2}\ell_{2}}^{\alpha\sigma}(r)
G_{\ell}^{\alpha}(r,r')
u_{f_{3}\ell_{3}}^{\alpha\sigma}(r')u_{f_{4}\ell_{4}}^{\alpha\sigma}(r')
r^{2}r'^{2}dr'dr +
Q_{\ell}
q_{\ell m}^{\alpha\sigma n\mathbf{k}n'\mathbf{k}'}
v_{n\mathbf{k}n'\mathbf{k}'}^{\alpha\sigma\ell m*}\left(R_{\alpha}\right)
\right],
\label{ExHFS}
\end{eqnarray}
\end{widetext}
where $Q_{\ell}=1/R_{ \alpha}^{\ell}$
and $q_{\ell m}^{\alpha\sigma n\mathbf{k}n'\mathbf{k}'}$ is Eq. (\ref{qlmnk})
for the unscreened potential or
$Q_{\ell}=\left(1/i_{\ell}\left(\lambda R_{{\alpha}}\right)\right)
\lambda^{\ell}/\left(2\ell+1\right)!!$
and $q_{\ell m}^{\alpha\sigma n\mathbf{k}n'\mathbf{k}'}$ is Eq. (\ref{qlmnks}) for
the screened potential.

As already mentioned above, the singularity which arises when
$\mathbf{q}=\mathbf{0}$
[see Eq. (\ref{vnkG})] needs to be considered properly.
Several methods to deal with this integrable singularity
when integrating into the Brillouin zone
are available in the literature
\cite{GygiPRB86,MassiddaPRB93,WenzienPRB95,PaierJCP05,CarrierPRB07,SpencerPRB08,FriedrichCPC09}
and have been used in very recent studies.
\cite{SorouriJCP06,NguyenPRB09,BroqvistPRB09,PaierPRB09,DucheminCPC10,HarlPRB10,BetzingerPRB10}
We adopted the simple scheme proposed by Spencer and Alavi
\cite{SpencerPRB08} which consists of multiplying Eq. (\ref{vnkG}) by
$1-\cos\left(\left\vert\mathbf{q}\right\vert R_{\text{c}}\right)$,
where $R_{\text{c}}=\left(3/\left(4\pi\right)N_{\mathbf{k}}\Omega\right)^{1/3}$
with $N_{\mathbf{k}}$ being the number of $\mathbf{k}$-points in the full
Brillouin zone. In the real space this corresponds to multiplying
Eq. (\ref{vc}) by the step function
$\theta(R_{\text{c}}-{\left\vert\mathbf{r}-\mathbf{r}'\right\vert})$.\cite{SpencerPRB08}
By doing this, the term $\mathbf{q}=\mathbf{0}$ tends to a finite value:
\begin{equation}
\lim_{\left\vert\mathbf{q}\right\vert\rightarrow0}\frac{4\pi}
{\left\vert\mathbf{q}\right\vert^{2}}
\left(1-\cos\left(\left\vert\mathbf{q}\right\vert R_{\text{c}}\right)\right) =
2\pi R_{\text{c}}^{2},
\label{limq0}
\end{equation}
which leads to a much more faster convergence (with respect to $N_{\mathbf{k}}$)
of the integrations into the Brillouin zone.

The screened potential has no singularity at $\mathbf{q}=\mathbf{0}$
[see Eq. (\ref{vnkGs})],
nevertheless it is still useful to apply the same technique in order
to accelerate further the convergence of the integrations into the Brillouin zone.
Multiplying Eq. (\ref{vcs}) by the step function
$\theta(R_{\text{c}}-{\left\vert\mathbf{r}-\mathbf{r}'\right\vert})$ means
that in the reciprocal space Eq. (\ref{vnkGs}) should be multiplied by
\begin{equation}
1-e^{-\lambda R_{\text{c}}}\left(\frac{\lambda}{\left\vert\mathbf{q}\right\vert}
\sin\left(\left\vert\mathbf{q}\right\vert R_{\text{c}}\right) +
\cos\left(\left\vert\mathbf{q}\right\vert R_{\text{c}}\right)\right),
\label{SAs}
\end{equation}
which becomes
$1-e^{-\lambda R_{\text{c}}}\left(\lambda R_{\text{c}} + 1\right)$ at
$\mathbf{q}=\mathbf{0}$.

\subsubsection{\label{VCCC}Valence-core and core-core terms}

By supposing that the core shells are closed (see Refs. \onlinecite{DagensPRB72}
and \onlinecite{MassiddaPRB93}), the Legendre polynomial addition
theorem\cite{Arfken} can be used to simplify
the calculation of the valence-core and core-core terms of the HF
exchange energy [Eqs. (\ref{ExvcHF}) and (\ref{ExccHF})].
The final expressions are given by
\begin{widetext}
\begin{eqnarray}
E_{\text{x,vc}}^{\text{HF}} & = &
-\sum_{\sigma}\sum_{\alpha}^{\text{cell}}
\sum_{n_{\text{c}},\ell_{\text{c}}}\sum_{n,\mathbf{k}}
\sum_{\ell,\ell',m'}\sum_{f_{1},f_{2}}
w_{n\mathbf{k}}^{\sigma}
\left(D_{\ell'm'}^{\alpha\sigma n\mathbf{k}f_{1}}\right)^{*}
D_{\ell'm'}^{\alpha\sigma n\mathbf{k}f_{2}}
C_{\ell0\ell_{\text{c}}0}^{\ell'0}
\sqrt{\frac{\left(2\ell_{\text{c}}+1\right)\left(2\ell+1\right)}
{4\pi\left(2\ell'+1\right)}} \nonumber\\
& & \times
\int\limits_{0}^{R_{\alpha}}\int\limits_{0}^{R_{\alpha}}
u_{f_{1}\ell'}^{\alpha\sigma}(r)
u_{n_{\text{c}}\ell_{\text{c}}}^{\alpha\sigma}(r)
H_{\ell}(r,r')
u_{n_{\text{c}}\ell_{\text{c}}}^{\alpha\sigma}(r')
u_{f_{2}\ell'}^{\alpha\sigma}(r')
r^{2}r'^{2}dr'dr,
\label{ExvcHF2}
\end{eqnarray}
\begin{eqnarray}
E_{\text{x,cc}}^{\text{HF}} & = &
-\frac{1}{2}\sum_{\sigma}\sum_{\alpha}^{\text{cell}}\sum_{n_{\text{c}},\ell_{\text{c}}
\atop n_{\text{c}}',\ell_{\text{c}}'}\sum_{\ell}
C_{\ell_{\text{c}}0\ell_{\text{c}}'0}^{\ell0}
\sqrt{\frac{\left(2\ell_{\text{c}}+1\right)\left(2\ell_{\text{c}}'+1\right)
\left(2\ell+1\right)}{4\pi}} \nonumber\\
& & \times
\int\limits_{0}^{R_{\alpha}}\int\limits_{0}^{R_{\alpha}}
u_{n_{\text{c}}'\ell_{\text{c}}'}^{\alpha\sigma}(r)
u_{n_{\text{c}}\ell_{\text{c}}}^{\alpha\sigma}(r)
H_{\ell}(r,r')
u_{n_{\text{c}}\ell_{\text{c}}}^{\alpha\sigma}(r')
u_{n_{\text{c}}'\ell_{\text{c}}'}^{\alpha\sigma}(r')
r^{2}r'^{2}dr'dr,
\label{ExccHF2}
\end{eqnarray}
\end{widetext}
where
$H_{\ell}(r,r')=\left(4\pi/\left(2\ell+1\right)\right)r_{<}^{\ell}/r_{>}^{\ell+1}$
for the unscreened potential or $H_{\ell}(r,r')=
4\pi\lambda i_{\ell}\left(\lambda r_{<}\right)k_{\ell}\left(\lambda r_{>}\right)$
for the screened potential.
$C_{\ell_{1}m_{1}\ell_{2}m_{2}}^{\ell_{3}m_{3}}$ are Gaunt coefficients
[Eq. (\ref{Gaunt})] and $D_{\ell m}^{\alpha\sigma n\mathbf{k}f}$ were defined
in Sec. \ref{Energy}. Note that in Eqs. (\ref{ExvcHF2}) and (\ref{ExccHF2}),
all integrations are inside the atomic spheres only, thus the cost for the
calculation of these two terms is negligible compared to the valence-valence
term.

\subsection{\label{Hamiltonian}Hamiltonian}

The HF exchange operator for the valence orbitals is the sum of the
valence-valence and valence-core terms:
$\hat{v}_{\text{x}\sigma}^{\text{HF}}=
\hat{v}_{\text{x}\sigma,\text{vv}}^{\text{HF}}+
\hat{v}_{\text{x}\sigma,\text{vc}}^{\text{HF}}$.
For the present work we chose to implement the HF
(and hybrid, see Sec. \ref{hybrid}) operator using a
second variational procedure, which consists of using
the semilocal (SL), LDA or GGA, orbitals as basis functions for the
calculation of the matrix elements of the perturbation operator
$\langle\psi_{n\mathbf{k}}^{\sigma\text{SL}}\vert
\hat{v}_{\text{x}\sigma}^{\text{HF}}-v_{\text{x}\sigma}^{\text{SL}}\vert
\psi_{n'\mathbf{k}}^{\sigma\text{SL}}\rangle$.
The HF part is given by
\begin{widetext}
\begin{equation}
\langle\psi_{n\mathbf{k}}^{\sigma\text{SL}}\vert\hat{v}_{\text{x}\sigma,\text{vv}}^{\text{HF}}\vert
\psi_{n'\mathbf{k}}^{\sigma\text{SL}}\rangle =
-\sum_{n'',\mathbf{k}''}w_{n''\mathbf{k}''}^{\sigma}
\int\limits_{\Omega}\int\limits_{\text{crystal}}
\psi_{n\mathbf{k}}^{\sigma\text{SL}*}(\mathbf{r})
\psi_{n''\mathbf{k}''}^{\sigma}(\mathbf{r})
v\left(\left\vert\mathbf{r}-\mathbf{r}'\right\vert\right)
\psi_{n''\mathbf{k}''}^{\sigma*}(\mathbf{r}')
\psi_{n'\mathbf{k}}^{\sigma\text{SL}}(\mathbf{r}')
d^{3}r'd^{3}r,
\label{psivxvvHFpsi}
\end{equation}
\begin{equation}
\langle\psi_{n\mathbf{k}}^{\sigma\text{SL}}\vert\hat{v}_{\text{x}\sigma,\text{vc}}^{\text{HF}}\vert
\psi_{n'\mathbf{k}}^{\sigma\text{SL}}\rangle =
-\sum_{\alpha}^{\text{cell}}\sum_{n_{\text{c}},\ell_{\text{c}},m_{\text{c}}}
\int\limits_{\text{S}_{\alpha}}\int\limits_{\text{S}_{\alpha}}
\psi_{n\mathbf{k}}^{\sigma\text{SL}*}(\mathbf{r})
\psi_{n_{\text{c}}\ell_{\text{c}}m_{\text{c}}}^{\alpha\sigma}(\mathbf{r})
v\left(\left\vert\mathbf{r}-\mathbf{r}'\right\vert\right)
\psi_{n_{\text{c}}\ell_{\text{c}}m_{\text{c}}}^{\alpha\sigma*}(\mathbf{r}')
\psi_{n'\mathbf{k}}^{\sigma\text{SL}}(\mathbf{r}')
d^{3}r'd^{3}r,
\label{psivxvcHFpsi}
\end{equation}
\end{widetext}
which are calculated using the same procedure as for the HF exchange energy, but with
$\rho_{n\mathbf{k}n''\mathbf{k}''}^{\sigma}=\psi_{n\mathbf{k}}^{\sigma\text{SL}*}
\psi_{n''\mathbf{k}''}^{\sigma}$ for Eq. (\ref{psivxvvHFpsi}).
The second variational procedure, which was also adopted for the HF
implementations in other LAPW codes
\cite{MassiddaPRB93,AsahiPRB99,AsahiPRB00,BetzingerPRB10}
leads to cheaper calculations, since in practice
the number of orbitals $\psi_{n\mathbf{k}}^{\sigma\text{SL}}$ which are
used for the construction of the HF Hamiltonian matrix is
much smaller than the number of LAPW basis functions.
In the present implementation, the core electrons experience the semilocal
potential, similarly as what is done in the \textsc{fleur} code, where
the core electrons are taken from a previous semilocal calculation and kept
frozen during the calculation with the hybrid functional.\cite{BetzingerPRB10}

\section{\label{hybrid}Screened hybrid functionals}

\begin{figure}
\includegraphics[scale=0.6]{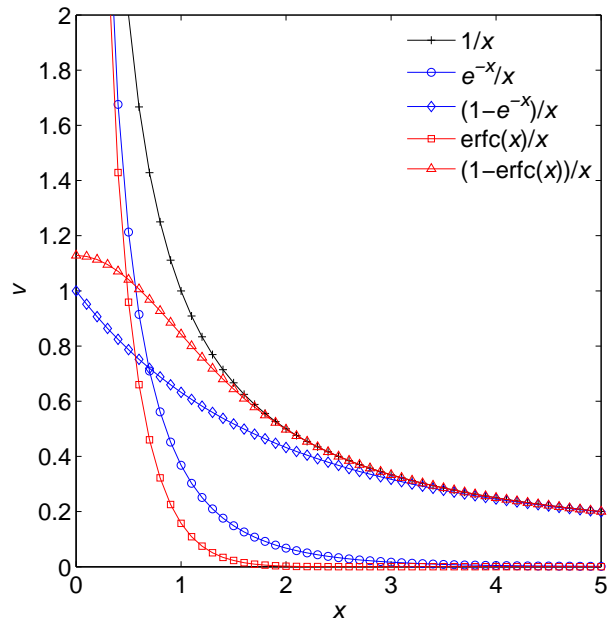}
\caption{\label{fig1}(Color online) Plots of the SR and LR parts of the Coulomb
operator $1/x$, when split using the exponential (in blue) or error
(in red) functions.}
\end{figure}

\begin{figure}
\includegraphics[scale=0.6]{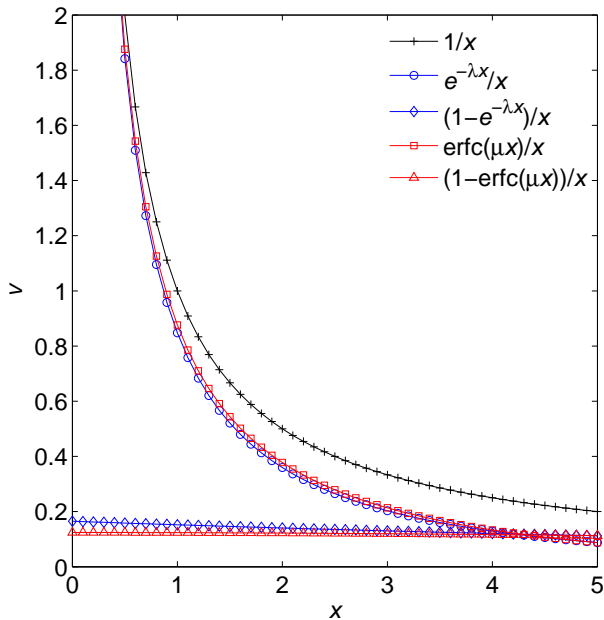}
\caption{\label{fig2}(Color online) Plots of the SR and LR parts of the Coulomb
operator $1/x$, when split using the error (with $\mu=0.11$ bohr$^{-1}$)
or the exponential [with $\lambda=\left(3/2\right)\mu=0.165$ bohr$^{-1}$]
functions.}
\end{figure}

In screened hybrid functionals, the SR part of a fraction $\alpha_{\text{x}}$ of
semilocal exchange is replaced
by SR HF exchange:\cite{HeydJCP03}
\begin{equation}
E_{\text{xc}} = E_{\text{xc}}^{\text{SL}} +
\alpha_{\text{x}}\left(E_{\text{x}}^{\text{SR-HF}} -
E_{\text{x}}^{\text{SR-SL}}\right),
\label{Exchybrid}
\end{equation}
where $E_{\text{x}}^{\text{SR-HF}}$ and $E_{\text{x}}^{\text{SR-SL}}$ are
obtained by replacing the full (i.e., unscreened) Coulomb operator by the
screened (i.e., SR) operator into the corresponding expressions.
For the HSE functional,\cite{HeydJCP03} the Coulomb operator was
split into SR and LR components by using the error function, however,
for the present work
we chose to split the Coulomb operator by using the exponential function:
\begin{equation}
\frac{1}{\left\vert\mathbf{r}-\mathbf{r}'\right\vert} =
\underbrace{
\frac{e^{-\lambda\left\vert\mathbf{r}-\mathbf{r}'\right\vert}}
{\left\vert\mathbf{r}-\mathbf{r}'\right\vert}}_{\text{SR}} +
\underbrace{
\frac{1-e^{-\lambda\left\vert\mathbf{r}-\mathbf{r}'\right\vert}}
{\left\vert\mathbf{r}-\mathbf{r}'\right\vert}}_{\text{LR}}.
\label{split}
\end{equation}
Figure \ref{fig1} shows the SR and LR parts [Eq. (\ref{split})] of the Coulomb
potential $1/x=1/\left\vert\mathbf{r}-\mathbf{r}'\right\vert$, and
for comparison, the same is shown when the error function is used
to split $1/x$ [$\text{erfc}(\mu x)=1-\text{erf}(\mu x)$ is the complementary
error function]. In both cases, the
screening parameter is set to $\lambda=\mu=1$. At $x=0$, the values of
the LR parts $\left(1-e^{-\lambda x}\right)/x$ and
$\left(1-\text{erfc}(\mu x)\right)/x$ are $\lambda$ and
$2\mu/\sqrt{\pi}$, respectively, thus these two ways of splitting the
Coulomb operator lead to LR components which are not zero at $x=0$.
Sharper splitting schemes which lead to a LR component which is zero at $x=0$
consist of using, e.g., the $\text{erfgau}$ function\cite{ToulousePRA04} or
simply the step function.\cite{SpencerPRB08}
We mention that for technical convenience, Shimazaki and Asai replaced
$e^{-\lambda x}$ by $\text{erfc}((2/3)\lambda x)$ in their proposed screened
HF potential.\cite{ShimazakiCPL08,ShimazakiJCP09,ShimazakiJCP10}
Indeed, from Fig. \ref{fig2} we can see that if
$\lambda=\left(3/2\right)\mu$, the two splitting procedures lead to very
similar SR and LR parts. In this example, $\mu=0.11$ bohr$^{-1}$, which is
the value used for the HSE06 functional.\cite{KrukauJCP06}

In Eq. (\ref{Exchybrid}), $E_{\text{x}}^{\text{SR-HF}}$ is given by
Eqs. (\ref{ExHF})-(\ref{ExccHF}) with
the Yukawa potential [Eq. (\ref{vcs})] for $v$ and
$E_{\text{x}}^{\text{SR-SL}}$ is given by
\begin{eqnarray}
E_{\text{x}}^{\text{SR-SL}} & = &
-\frac{3}{4}\left(\frac{6}{\pi}\right)^{1/3}
\sum_{\sigma}\int\limits_{\Omega}\rho_{\sigma}^{4/3}(\mathbf{r}) \nonumber \\
& & \times F_{\text{x}}(s_{\sigma}(\mathbf{r}))J(a_{\sigma}(\mathbf{r}))d^{3}r,
\label{ExSRSL}
\end{eqnarray}
where $F_{\text{x}}(s_{\sigma})$
[where $s_{\sigma}=\left\vert\nabla\rho_{\sigma}\right\vert/
\left(2\rho_{\sigma}k_{\text{F}}^{\sigma}\right)$ with
$k_{\text{F}}^{\sigma}=\left(6\pi^{2}\rho_{\sigma}\right)^{1/3}$]
is the enhancement factor of the semilocal exchange
functional and $J(a_{\sigma})$
[where
$a_{\sigma}=\lambda\sqrt{F_{\text{x}}(s_{\sigma})}/\left(2k_{\text{F}}^{\sigma}\right)$]
is a function, whose analytical form depends on the way the Coulomb operator
is screened [$J(a_{\sigma})=1$ for the unscreened operator].
In our case [SR part of Eq. (\ref{split})], $J(a_{\sigma})$ is given by
\cite{RobinsonPRL62}
\begin{eqnarray}
J(a_{\sigma}) & = & 1 - \frac{2}{3}a_{\sigma}^{2} -
\frac{8}{3}a_{\sigma}\arctan\frac{1}{a_{\sigma}} \nonumber \\
& & + \frac{2}{3}a_{\sigma}^{2}\left(a_{\sigma}^{2}+3\right)
\ln\left(1+\frac{1}{a_{\sigma}^{2}}\right).
\label{J}
\end{eqnarray}
Equation (\ref{ExSRSL})
is an approximation which was originally proposed by Iikura
\textit{et al}.,\cite{IikuraJCP01} but with the function $J(a_{\sigma})$
for the error function.
Recently, Akinaga and Ten-no \cite{AkinagaCPL08} used Eq. (\ref{ExSRSL})
in conjunction with the Yukawa potential as in
the present work. (However, we note that in Refs. \onlinecite{IikuraJCP01} and
\onlinecite{AkinagaCPL08}, this is the LR part of the semilocal exchange
which was replaced by LR HF.) A more elegent way of
calculating $E_{\text{x}}^{\text{SR-SL}}$ would be to use its
expression in terms of the exchange hole (as done for HSE
\cite{ErnzerhofJCP98,HeydJCP04a}), and, for practical convenience,
to find a mathematical form for the exchange hole such that an analytical
integration with the Yukawa potential is possible, as done in
Ref. \onlinecite{HendersonJCP08} for the error function.
This method has been used in Ref. \onlinecite{SchimkaJCP11} for the HSEsol
functional, which is based on the PBEsol GGA functional.\cite{PerdewPRL08}
We did not consider this possibility for the present work.

For the semilocal terms in Eq. (\ref{Exchybrid}) we have chosen
PBE,\cite{PerdewPRL96} which is of the GGA form.
In the following, this functional will be called
YS-PBE0 (where YS stands for Yukawa screened).
In the literature, the unscreened version of this hybrid functional
(recovered when $\lambda\rightarrow0$)
is called PBE0,\cite{ErnzerhofJCP99,AdamoJCP99} for which the
fraction of HF exchange is
$\alpha_{\text{x}}=0.25$ (see Ref. \onlinecite{PerdewJCP96}).
For $\lambda\rightarrow\infty$, YS-PBE0 reduces to PBE.
The calculation of the total energy for hybrid functionals is given in
Appendix \ref{totalenergy}.
As already mentioned in Sec. \ref{Hamiltonian}, the second variational procedure
has been implemented, and the matrix elements of the perturbation operator
corresponding to Eq. (\ref{Exchybrid}) are given by
\begin{equation}
\langle\psi_{n\mathbf{k}}^{\sigma\text{SL}}\vert
\alpha_{\text{x}}\left(\hat{v}_{\text{x}\sigma}^{\text{SR-HF}}-
v_{\text{x}\sigma}^{\text{SR-SL}}\right)\vert
\psi_{n'\mathbf{k}}^{\sigma\text{SL}}\rangle,
\label{psivxHFvxSRSLpsi}
\end{equation}
where the expression for $v_{\text{x}\sigma}^{\text{SR-SL}}=
\delta E_{\text{x}}^{\text{SR-SL}}/\delta\rho_{\sigma}$ is given
in Appendix \ref{Functionalderivative}.

\section{\label{results}Numerical results}

The calculations presented in this section were done with
values for the parameters such that the results are well converged.
The most important parameters are the number of $\mathbf{k}$ points for
the integrations into the Brillouin zone, the size of the basis sets
(first and second variational procedures), $G_{\text{max}}$ and
$\ell_{\text{max}}$ in Eqs. (\ref{rhonk2}) and (\ref{vnk2}), and
$\ell_{\text{max}}$ in Eq. (\ref{rhonklm}). We will not discuss in detail the
convergence of the results with respect to these parameters, but just mention
the following: for the transition energies and lattice constants, the number of
orbitals used as basis functions for the second variational procedure is
between two and six times larger than the number of valence bands in the system.
The values of $G_{\text{max}}$ lie in the range 4$-$10 bohr$^{-1}$ and for most
calculations the value $\ell_{\text{max}}=4$ was used, which is more than enough
most of the time. The size of the $\mathbf{k}$-meshes will be mentioned below.

We mention again that the computation of the HF Hamiltonian is very
expensive, and for the systems we have considered
this leads to computational times which are by one or two orders of magnitude larger
than for semilocal functionals. Actually, the values of all parameters mentioned above
have a large impact on the computational time.

\subsection{\label{Comparison}Comparison with other codes}

\subsubsection{\label{HF}HF energy}

\begin{table}
\caption{\label{table1}Total and exchange energies (in Ha) of He atom.}
\begin{ruledtabular}
\begin{tabular}{lcccc}
\multicolumn{1}{l}{} &
\multicolumn{2}{c}{\textsc{wien2k}} &
\multicolumn{2}{c}{Reference} \\
\cline{2-3}\cline{4-5}
\multicolumn{1}{l}{Functional} &
\multicolumn{1}{c}{$-E_{\text{tot}}$} &
\multicolumn{1}{c}{$-E_{\text{x}}$} &
\multicolumn{1}{c}{$-E_{\text{tot}}$} &
\multicolumn{1}{c}{$-E_{\text{x}}$} \\
\hline
LDAx\footnotemark[1]  & 2.724 & 0.853 & 2.724 & 0.853 \\
B88\footnotemark[1]   & 2.863 & 1.016 & 2.863 & 1.016 \\
PW91x\footnotemark[1] & 2.855 & 1.005 & 2.855 & 1.005 \\
HF\footnotemark[1]    & 2.862 & 1.024 & 2.862 & 1.026 \\
HF\footnotemark[2]    &       & 0.998 &       & 0.998 \\
HF\footnotemark[3]    &       & 1.017 &       &       \\
\end{tabular}
\end{ruledtabular}
\footnotetext[1]{Obtained from exchange-only self-consistent calculations.
The reference results are from Refs. \onlinecite{ZhuJCP93} and
\onlinecite{FroeseFischer}.}
\footnotetext[2]{Evaluated with LDA (exchange and correlation) orbitals.
The reference result is from Ref. \onlinecite{BeckeJCP07}.}
\footnotetext[3]{Evaluated with B88PW91 orbitals.}
\end{table}

As a first test of the correctness and accuracy of the implementation,
we considered systems which do not contain core electrons, such that all
electrons are treated self-consistently with the HF method.
The He atom and solid LiH are two such systems for which
highly accurate HF results are available in the literature.
The LDAx (exchange-only LDA) orbitals were used as basis functions for
the Hamiltonian of the second variational procedure.

\begin{figure}
\includegraphics[scale=0.6]{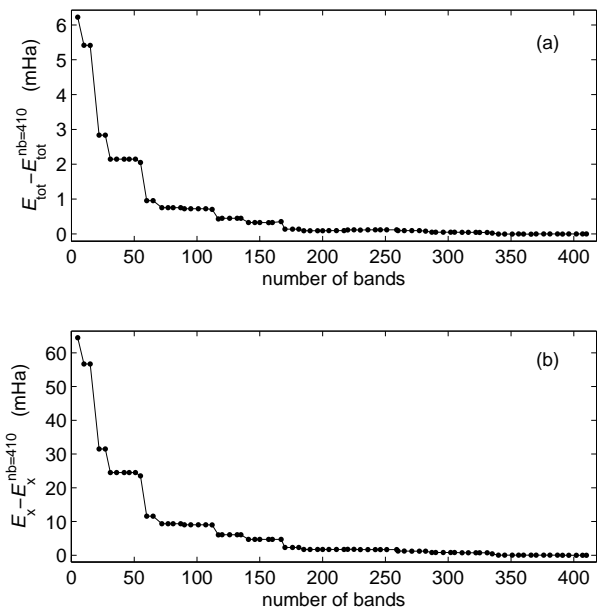}
\caption{\label{fig3} (a) Total and (b) exchange energy of the He atom
with respect to the values calculated with 410 bands.}
\end{figure}

The results for the He atom are shown in Table \ref{table1}.
The calculations were done in a fcc cell with a lattice constant of
9.5 \AA~which is large enough to make the interactions between the He atoms
negligible.
First, in order to have an idea of the accuracy that can be expected with
\textsc{wien2k}, we did calculations with semilocal functionals
(exchange only: LDAx, B88,\cite{BeckePRA88} and PW91x\cite{PerdewPRB92b})
and compared them to accurate atomic results.\cite{ZhuJCP93,BeckeJCP07}
From the results we can see that an agreement at the mHa level can be reached,
which is the target for the HF calculations.
The self-consistent HF results shown in Table \ref{table1} were obtained
using 410 bands for the second variational procedure, which was enough to
reach convergence and thus agreement with accurate atomic results.\cite{FroeseFischer}
However, for the exchange energy $E_{\text{x}}$, the agreement with the
reference result is not perfect. Actually, we can see in Fig. \ref{fig3}
that for a given number of bands, the error with respect to the
(approximately) converged value is ten times larger for the exchange energy
than for the total energy. For $E_{\text{tot}}$, about 120 bands are necessary
to reach convergence at the mHa level, while 410
bands are still not enough for $E_{\text{x}}$
(about 3000 LAPW basis functions are used for the first variational procedure).
It is known that the total energy converges faster than its components.
In order to evaluate the effects due to self-consistency, the HF exchange energy
was also evaluated using the LDA (with PW92 for correlation\cite{PerdewPRB92a})
and B88PW91 (B88\cite{BeckePRA88} for exchange and PW91\cite{PerdewPRB92b}
for correlation) orbitals.
From Table \ref{table1}, we can see that using LDA orbitals leads to
an HF exchange energy whose magnitude is 26 mHa smaller, while using
B88PW91 orbitals leads to a value which is much closer to the
self-consistent one, which is maybe not surprising since the empirical
parameter in B88 was determined by a fit to HF exchange energy of
rare-gas atoms.\cite{BeckePRA88}
This indicates that using the B88PW91 orbitals as basis functions for the
second-variational Hamiltonian would be more efficient.

In Ref. \onlinecite{PaierPRB09} (as well as in the Comment),
well converged calculations on solid LiH (rocksalt structure) using Gaussian
basis sets yield a value of $E_{\text{tot}}=-8.0645$ Ha at the experimental geometry
(4.084 \AA). Using a $6\times6\times6$ $\mathbf{k}$-mesh and 65 bands for the second
variational procedure we obtained $E_{\text{tot}}=-8.0642$ Ha. Increasing further the
number of bands would lower the total energy and reduce the difference between the
Gaussian and LAPW results. Therefore, as for the He atom, the HF energy calculated
with the LAPW code agrees very well with the literature results.

\subsubsection{\label{PBE0}PBE0 calculations}

\begin{table*}
\caption{\label{table2}Transition energies (in eV) obtained with the PBE, PBE0,
and YS-PBE0 ($\lambda=0.165$ bohr$^{-1}$) functionals.}
\begin{ruledtabular}
\begin{tabular}{llccccccccc}
\multicolumn{1}{l}{} &
\multicolumn{1}{l}{} &
\multicolumn{3}{c}{\textsc{wien2k}} &
\multicolumn{3}{c}{\textsc{vasp}\footnotemark[1]} &
\multicolumn{2}{c}{\textsc{fleur}\footnotemark[2]} \\
\cline{3-5}\cline{6-8}\cline{9-10}
\multicolumn{1}{l}{Solid} &
\multicolumn{1}{c}{Transition} &
\multicolumn{1}{c}{PBE} &
\multicolumn{1}{c}{PBE0} &
\multicolumn{1}{c}{YS-PBE0} &
\multicolumn{1}{c}{PBE} &
\multicolumn{1}{c}{PBE0} &
\multicolumn{1}{c}{HSE06} &
\multicolumn{1}{c}{PBE} &
\multicolumn{1}{c}{PBE0} &
\multicolumn{1}{c}{Expt.\footnotemark[3]} \\
\hline
Ar   & $\Gamma\rightarrow\Gamma$ & 8.69 & 11.09 & 10.36 & 8.68 & 11.09 & 10.34 & 8.71 & 11.15 & 14.2        \\
C    & $\Gamma\rightarrow\Gamma$ & 5.59 &  7.69 &  6.94 & 5.59 &  7.69 &  6.97 & 5.64 &  7.74 &  7.3        \\
     & $\Gamma\rightarrow X$     & 4.76 &  6.64 &  5.91 & 4.76 &  6.66 &  5.91 & 4.79 &  6.69 &             \\
     & $\Gamma\rightarrow L$     & 8.46 & 10.76 &  9.97 & 8.46 & 10.77 & 10.02 & 8.58 & 10.88 &             \\
Si   & $\Gamma\rightarrow\Gamma$ & 2.56 &  3.95 &  3.30 & 2.57 &  3.97 &  3.32 & 2.56 &  3.96 &  3.4        \\
     & $\Gamma\rightarrow X$     & 0.71 &  1.91 &  1.31 & 0.71 &  1.93 &  1.29 & 0.71 &  1.93 &             \\
     & $\Gamma\rightarrow L$     & 1.53 &  2.86 &  2.23 & 1.54 &  2.88 &  2.24 & 1.54 &  2.87 &  2.4        \\
GaAs & $\Gamma\rightarrow\Gamma$ & 0.53 &  1.99 &  1.39 & 0.56 &  2.01 &  1.45 & 0.55 &  2.02 &  1.63       \\
     & $\Gamma\rightarrow X$     & 1.46 &  2.66 &  2.08 & 1.46 &  2.67 &  2.02 & 1.47 &  2.69 &  2.18, 2.01 \\
     & $\Gamma\rightarrow L$     & 1.01 &  2.35 &  1.74 & 1.02 &  2.37 &  1.76 & 1.02 &  2.38 &  1.84, 1.85 \\
MgO  & $\Gamma\rightarrow\Gamma$ & 4.79 &  7.23 &  6.49 & 4.75 &  7.24 &  6.50 & 4.84 &  7.31 &  7.7        \\
     & $\Gamma\rightarrow X$     & 9.16 & 11.58 & 10.83 & 9.15 & 11.67 & 10.92 & 9.15 & 11.63 &             \\
     & $\Gamma\rightarrow L$     & 7.95 & 10.43 &  9.68 & 7.91 & 10.38 &  9.64 & 8.01 & 10.51 &             \\
NaCl & $\Gamma\rightarrow\Gamma$ & 5.22 &  7.29 &  6.61 & 5.20 &  7.26 &  6.55 & 5.08 &  7.13 &  8.5        \\
     & $\Gamma\rightarrow X$     & 7.59 &  9.80 &  9.06 & 7.60 &  9.66 &  8.95 & 7.39 &  9.59 &             \\
     & $\Gamma\rightarrow L$     & 7.33 &  9.40 &  8.70 & 7.32 &  9.41 &  8.67 & 7.29 &  9.33 &             \\
\end{tabular}
\end{ruledtabular}
\footnotetext[1]{Reference \onlinecite{PaierJCP06} (see Erratum for HSE06 results).}
\footnotetext[2]{Reference \onlinecite{BetzingerPRB10}.}
\footnotetext[3]{The references for the experimental values are given in
Table I of Ref. \onlinecite{BetzingerPRB10}.}
\end{table*}

In Refs. \onlinecite{PaierJCP06} and \onlinecite{BetzingerPRB10}, calculations
with the unscreened hybrid functional PBE0
were done within the projector augmented-wave
(\textsc{vasp} code) and LAPW (\textsc{fleur} code) methods,
respectively. The implementation of the HF equations within the LAPW basis set
as reported in Ref. \onlinecite{BetzingerPRB10} was done using another technique
(mixed product basis) as the one used in the present work (pseudocharge method).
The integrations into the Brillouin zone were done with a
$7\times7\times7$ $\mathbf{k}$-mesh for the semiconductors and insulators,
while for the metals Li, Cu, and Rh a $12\times12\times12$ $\mathbf{k}$-mesh was used.
The results from Refs. \onlinecite{PaierJCP06} and \onlinecite{BetzingerPRB10}
were done with a $12\times12\times12$ $\mathbf{k}$-mesh, however test calculations
indicate that our results are converged within $\sim0.02$ eV for the transition
energies and $\sim0.002$ \AA~for the lattice constants.

Transition energies were calculated for six solids at the
experimental lattice constant:
Ar (fcc, 5.260 \AA), C (diamond, 3.567 \AA), Si (diamond, 5.430 \AA),
GaAs (zinc blende, 5.648 \AA), MgO (rocksalt, 4.207 \AA), and
NaCl (rocksalt, 5.595 \AA).
The PBE0 results, as well as the PBE and experimental results, are given
in Table \ref{table2}, where we can see that
the \textsc{wien2k} results agree very well with the
\textsc{fleur} and \textsc{vasp} results. There are a few cases
where the discrepancy is larger than 0.1 eV.
For the $\Gamma\rightarrow L$ transition in C, there is a difference
of 0.12 eV between the \textsc{wien2k} and \textsc{fleur} values and
in the case of NaCl, a disagreement of 0.15$-$0.2 eV with \textsc{fleur} is
found for the $\Gamma\rightarrow\Gamma$ and $\Gamma\rightarrow X$ transitions.
Nevertheless, overall the agreement with the \textsc{fleur} and \textsc{vasp}
codes for the PBE0 hybrid functional is clearly satisfactory, in particular
with \textsc{vasp}.
Compared to the experimental values, the PBE0 functional clearly improves
upon PBE, however, some sizeable disagreements with experiment are still
present, as for example for Ar and NaCl for which PBE0 underestimates the
$\Gamma\rightarrow\Gamma$ transition by about 3 and 1.2 eV, respectively.
In general, the tendency of the PBE0 functional is to overestimate
small band gaps (e.g., GaAs) and to underestimate large band gaps
(e.g., rare-gas solids).\cite{MarquesPRB11}

\begin{table*}
\caption{\label{table3}Equilibrium lattice constants $a_{0}$ (in \AA) and bulk
moduli $B_{0}$ (in GPa) obtained with the PBE, PBE0, and YS-PBE0
($\lambda=0.165$ bohr$^{-1}$) functionals. The experimental values, which are corrected
for the zero-point anharmonic expansion, are from Ref. \onlinecite{SchimkaJCP11}.}
\begin{ruledtabular}
\begin{tabular}{lcccccccccccccc}
\multicolumn{1}{l}{} &
\multicolumn{6}{c}{\textsc{wien2k}} &
\multicolumn{6}{c}{\textsc{vasp}\footnotemark[1]} &
\multicolumn{2}{c}{} \\
\cline{2-7}\cline{8-13}
\multicolumn{1}{l}{} &
\multicolumn{2}{c}{PBE} &
\multicolumn{2}{c}{PBE0} &
\multicolumn{2}{c}{YS-PBE0} &
\multicolumn{2}{c}{PBE} &
\multicolumn{2}{c}{PBE0} &
\multicolumn{2}{c}{HSE06} &
\multicolumn{2}{c}{Expt.} \\
\cline{2-3}\cline{4-5}\cline{6-7}\cline{8-9}\cline{10-11}\cline{12-13}\cline{14-15}
\multicolumn{1}{l}{Solid} &
\multicolumn{1}{c}{$a_{0}$} &
\multicolumn{1}{c}{$B_{0}$} &
\multicolumn{1}{c}{$a_{0}$} &
\multicolumn{1}{c}{$B_{0}$} &
\multicolumn{1}{c}{$a_{0}$} &
\multicolumn{1}{c}{$B_{0}$} &
\multicolumn{1}{c}{$a_{0}$} &
\multicolumn{1}{c}{$B_{0}$} &
\multicolumn{1}{c}{$a_{0}$} &
\multicolumn{1}{c}{$B_{0}$} &
\multicolumn{1}{c}{$a_{0}$} &
\multicolumn{1}{c}{$B_{0}$} &
\multicolumn{1}{c}{$a_{0}$} &
\multicolumn{1}{c}{$B_{0}$} \\
\hline
Li  & 3.434 &  13.9 & 3.464 &  13.1 & 3.467 &  12.6 & 3.438 &  13.7 & 3.463 &  13.7 & 3.460 &  13.6 & 3.453 &  13.9 \\
C   & 3.575 & 435   & 3.549 & 475   & 3.554 & 467   & 3.574 & 431   & 3.549 & 467   & 3.549 & 467   & 3.553 & 455   \\
Si  & 5.476 &  89.0 & 5.443 &  99.4 & 5.459 &  96.5 & 5.469 &  87.8 & 5.433 &  99.0 & 5.435 &  97.7 & 5.421 & 101   \\
Cu  & 3.631 & 141   & 3.630 & 131   & 3.654 & 119   & 3.635 & 136   & 3.636 & 130   & 3.638 & 126   & 3.595 & 145   \\
Rh  & 3.830 & 256   & 3.787 & 292   & 3.799 & 280   & 3.830 & 254   & 3.785 & 291   & 3.783 & 288   & 3.794 & 272   \\
LiF & 4.069 &  67.2 & 4.008 &  70.2 & 4.035 &  67.3 & 4.068 &  67.3 & 4.011 &  72.8 & 4.018 &  72.7 & 3.972 &  76.3 \\
BN  & 3.628 & 374   & 3.601 & 407   & 3.607 & 401   & 3.626 & 370   & 3.600 & 402   & 3.600 & 402   & 3.592 & 410   \\
SiC & 4.384 & 213   & 4.352 & 242   & 4.361 & 236   & 4.380 & 210   & 4.347 & 231   & 4.348 & 230   & 4.346 & 229   \\
\end{tabular}
\end{ruledtabular}
\footnotetext[1]{Reference \onlinecite{PaierJCP06} (see Erratum for HSE06 results).}
\end{table*}

The lattice constant and bulk modulus of a few selected compounds,
namely, Li (bcc), C (diamond), Si (diamond), Cu (fcc), Rh (fcc),
LiF (rocksalt), BN (zinc blende),
and SiC (zinc blende) were calculated using the PBE and PBE0 functionals.
The results are shown in Table \ref{table3} together with the values obtained
with the \textsc{vasp} code \cite{PaierJCP06} and the experimental data,
which were corrected for the zero-point anharmonic expansion.\cite{SchimkaJCP11}
By comparing the \textsc{wien2k} and \textsc{vasp} results, we can see that
excellent agreement between the two codes are obtained both for the PBE and PBE0
functionals. The largest discrepancy in
the lattice constant is found for Si, where a difference of 0.007$-$0.01~\AA~
is found for PBE and PBE0.
From Table \ref{table3} we can see that
there is also a good agreement between the two codes for the bulk
modulus. On average, the hybrid functional PBE0 improves
over the GGA PBE for the lattice constant and bulk modulus of
semiconductors and metals as shown in Ref. \onlinecite{PaierJCP06}.

\subsubsection{\label{YSPBE}YS-PBE0 calculations}

As mentioned in Sec. \ref{hybrid} (see Fig. \ref{fig2}), choosing
$\lambda=\left(3/2\right)\mu$ in Eq. (\ref{split}) leads to a splitting
of the Coulomb operator which is very similar to the one obtained by using
the error function with a given $\mu$.
\cite{ShimazakiCPL08,ShimazakiJCP09,ShimazakiJCP10}
In the HSE06 functional,\cite{KrukauJCP06}
$\mu$ is fixed to 0.11 bohr$^{-1}$ and in order to see how well the
YS-PBE0 functional can reproduce the HSE06 transition energies (see Erratum of
Ref. \onlinecite{PaierJCP06}), calculations with
$\lambda=\left(3/2\right)0.11=0.165$ bohr$^{-1}$ were done.
From the results shown in Table \ref{table2}, we can see that the agreement
between HSE06 (\textsc{vasp}) and YS-PBE0 is as good as it was for PBE0 with
differences smaller than 0.03 eV in most cases.
Compared to PBE0, the screened hybrid functionals lead to better (worse)
agreement with experiment for small (large) band gaps (see also
Ref. \onlinecite{MarquesPRB11}).

The YS-PBE0 results for the lattice constant and bulk modulus are shown in
Table \ref{table3}. The agreement between the HSE06 and YS-PBE0 results is
fairly good in cases like Li or C, while larger differences
can be seen for Si (0.024 \AA), LiF (0.017 \AA), Cu (0.016 \AA),
and Rh (0.016 \AA). An important
contribution to these differences in the lattice constant between the
HSE06 and YS-PBE0 values could be attributed to the different schemes used for the
screening of the semilocal exchange term [Eq. (\ref{ExSRSL})]. For YS-PBE0,
the method of Iikura \textit{et al}.\cite{IikuraJCP01} is used,
while in HSE06 the screened exchange energy is obtained by integrating
a model of the exchange hole.\cite{HeydJCP04a,ErnzerhofJCP98} However, it
seems that using one of the scheme or the other has very little influence on
the transition energies as shown above.

\subsection{\label{Cu2O}Cu$_{2}$O}

\begin{table}
\caption{\label{table4}Band gap (in eV) and Cu EFG (in $10^{21}$ V/m$^{2}$)
of Cu$_{2}$O calculated at the experimental lattice constant (4.27~\AA).}
\begin{tabular}{lcccc}
\hline
\hline
\multicolumn{1}{l}{} &
\multicolumn{1}{c}{} &
\multicolumn{3}{c}{EFG} \\
\cline{3-5}
\multicolumn{1}{l}{Method} &
\multicolumn{1}{c}{Band gap} &
\multicolumn{1}{c}{Total} &
\multicolumn{1}{c}{$p$-$p$} &
\multicolumn{1}{c}{$d$-$d$} \\
\hline
LDA                                             & 0.53                 & $-5.3$             & $-16.0$ & 10.5 \\
PBE                                             & 0.53                 & $-5.5$             & $-16.4$ & 10.6 \\
B88PW91                                         & 0.55                 & $-5.6$             & $-16.4$ & 10.6 \\
EV93PW91                                        & 0.57                 & $-6.6$             & $-17.4$ & 10.6 \\
LDA+$U$ (FLL, $U=4$ eV)                         & 0.65                 & $-6.1$             & $-16.1$ &  9.8 \\
LDA+$U$ (FLL, $U=8$ eV)                         & 0.80                 & $-6.6$             & $-16.4$ &  9.5 \\
LDA+$U$ (FLL, $U=12$ eV)                        & 0.91                 & $-7.6$             & $-16.5$ &  8.8 \\
LDA+$U$ (AMF, $U=4$ eV)                         & 0.63                 & $-4.8$             & $-16.0$ & 11.0 \\
LDA+$U$ (AMF, $U=8$ eV)                         & 0.79                 & $-2.6$             & $-16.2$ & 13.4 \\
LDA+$U$ (AMF, $U=12$ eV)                        & 0.94                 &   0.6              & $-16.6$ & 17.0 \\
PBE0 (onsite)                                   & 0.79                 & $-3.4$             & $-16.4$ & 12.8 \\
PBE0                                            & 2.77                 & $-8.5$             & $-19.5$ & 10.8 \\
YS-PBE0                                         & 1.99                 & $-8.3$             & $-19.3$ & 10.8 \\
pseudo-SIC\footnotemark[1]                      & 1.80 \\
B3LYP\footnotemark[2]                           & 2.1  \\
HSE ($\alpha_{\text{x}}=0.275$)\footnotemark[3] & 2.12 \\
sc$GW$\footnotemark[4]                          & 1.97 \\
Expt.                                           & 2.17\footnotemark[5] & 9.8\footnotemark[6] \\
\hline
\hline
\end{tabular}
\footnotetext[1]{Reference \onlinecite{FilippettiPRB05}.}
\footnotetext[2]{Reference \onlinecite{HuPRB08}.}
\footnotetext[3]{Reference \onlinecite{ScanlonJPCL10}.}
\footnotetext[4]{References \onlinecite{BrunevalPRL06}.}
\footnotetext[5]{References \onlinecite{BaumeisterPR61}.}
\footnotetext[6]{Only the magnitude is known. Calculated using
$Q\left(^{63}\text{Cu}\right)=0.22$.\cite{KushidaPR56,PyykkoMP01}}
\end{table}

\begin{figure}
\includegraphics[scale=0.50]{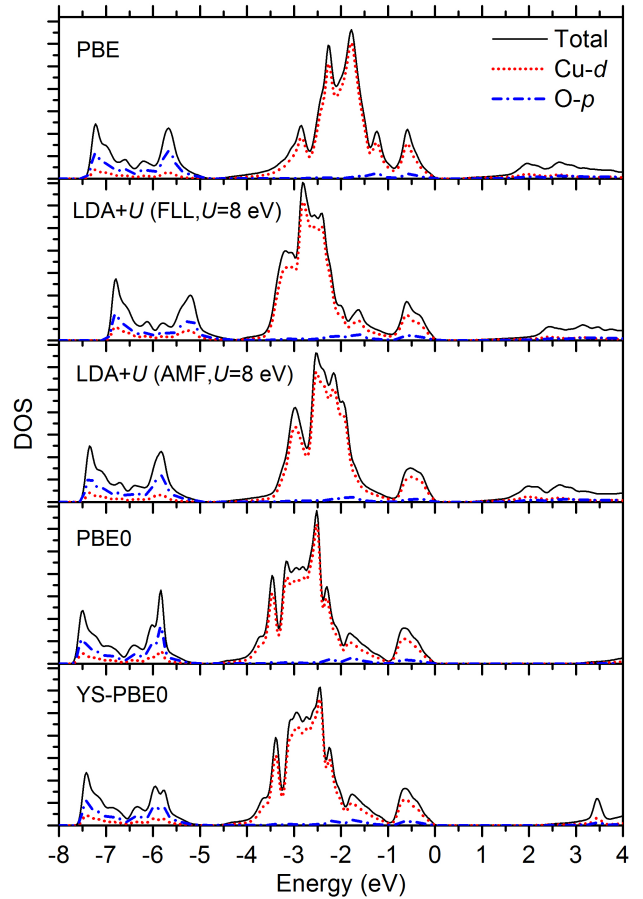}
\caption{\label{fig4}(Color online) Density of states of Cu$_{2}$O calculated
with different functionals. The Fermi energy is set at $E=0$ eV.}
\end{figure}

Cuprous oxide (Cu$_{2}$O) is a semiconductor which has been used in many
applications (e.g., catalysis and photovoltaics). Its structure is cubic
(space group $Pn\overline{3}m$) and the unit cell, which has a lattice
constant of 4.27~\AA,\cite{WernerPRB82} contains six atoms. In this structure,
shown in Fig. 1 of Ref. \onlinecite{MarksteinerZPB86}, the O atoms are fourfold
coordinated by Cu atoms, whereas the Cu atoms are linearly coordinated by O
atoms. Formally Cu has a valency of $+1$ and
the Cu-$3d$ shell in Cu$_{2}$O is full, therefore the correlation effects
in the Cu-$3d$ shell should not play an important role as it is the
case for CuO.\cite{GhijsenPRB08}

Many experimental and theoretical studies on Cu$_{2}$O have been done.
On the theoretical side it has been shown that the semilocal approximations
underestimate the band gap as expected (see Refs. \onlinecite{HuPRB08} and
\onlinecite{ScanlonJCP09} for collections of previously done calculations),
but also the Cu electric-field gradient (EFG),\cite{LaskowskiPRB03} which is
a ground-state property derived from the electron density.
LDA+$U$ (or GGA+$U$) improves
only slightly over the semilocal approximations,
\cite{ScanlonJCP09,LaskowskiPRB03} while the pseudo self-interaction method
(pseudo-SIC),\cite{FilippettiPRB05} the hybrid functionals,
\cite{HuPRB08,ScanlonPRL09,ScanlonJPCL10} and self-consistent $GW$
(sc$GW$)\cite{BrunevalPRL06} provide band gaps in much better agreement with
experiment. Actually, the results for the EFG show that the semilocal and
LDA+$U$ methods do not provide an accurate description of the occupied states.

In Table \ref{table4}, we show the results for the band gap and EFG obtained
with the hybrid functionals PBE0 and YS-PBE0 ($\lambda=0.165$ bohr$^{-1}$), which
were obtained with a mesh of $5\times5\times5$ $\mathbf{k}$-points.
The calculations with the semilocal (LDA,\cite{PerdewPRB92a} PBE,\cite{PerdewPRL96}
B88PW91,\cite{BeckePRA88,PerdewPRB92b} and EV93PW91\cite{EngelPRB93,PerdewPRB92b}),
LDA+$U$ [fully localized limit (FLL)\cite{CzyzykPRB94} and around mean-field
(AMF)\cite{CzyzykPRB94} versions],
and onsite PBE0 \cite{TranPRB06} methods (results in Table \ref{table4}) were done
with a $12\times12\times12$ $\mathbf{k}$-mesh.
The radii of the Cu and O atomic spheres are 1.84 and 1.63 bohr, respectively.
LDA, PBE, and B88PW91 give values for the band gap
($\sim0.5$ eV) and EFG ($\sim-5.5\times10^{21}$ V/m$^{2}$)
which are much smaller than the experimental values (above 2 eV for the band
gap\cite{BaumeisterPR61} and $9.8\times10^{21}$ V/m$^{2}$ for the
EFG\cite{KushidaPR56,PyykkoMP01}).
EV93PW91 improves for the EFG with a value of $-6.6\times10^{21}$ V/m$^{2}$, but
not for the band gap contrary to what was reported for many other solids in
Ref. \onlinecite{TranJPCM07}.
LDA+$U$ slightly improves the results for the band gap and
its two versions, FLL and AMF, lead to the same value for a given
value of the Coulomb parameter $U$ (the exchange parameter $J$ has been fixed to
0.95 eV). However, this improvement is minor and even with $U=12$ eV the band gap
remains well below the experimental value. For the EFG, FLL and AMF lead to different
trends. An increase of $U$ leads to an increase of the magnitude of the EFG with FLL,
while the opposite is obtained with AMF, which yields a positive value for $U=12$ eV.
In Table \ref{table4}, the $p$-$p$ and $d$-$d$ components (inside the Cu atomic sphere)
of the EFG are also shown. As expected, the change in the EFG due to $U$ comes mainly
from the $d$-$d$ part.
The onsite PBE0 method slightly improves the results for the band gap (0.8 eV),
but significantly decreases the EFG ($-3.4\times10^{21}$ V/m$^{2}$).
Overall, the FLL version of LDA+$U$ leads to a moderate improvement over the
semilocal functionals, while AMF and onsite PBE0 behave similarly by reducing
the magnitude of the EFG.

The results obtained with PBE0 and YS-PBE0 are in much better agreement with
experiment. In particular, the screened YS-PBE0 functional leads to a band gap
of 1.99 eV, which is very close to the experimental value, and an
EFG of $-8.3\times10^{21}$ V/m$^{2}$ which is much closer to experiment compared
to the values obtained with other functionals. PBE0 leads to a band gap which
seems to be too high and an EFG very similar to YS-PBE0.
Focusing now on the PBE and PBE0 results for the EFG,
we can see from the decomposition of the EFG (Table \ref{table4}) that
the increase in magnitude of the EFG by going from PBE to PBE0 comes
mainly from the $p$-$p$ component. By decomposing further the $p$-$p$
component, we could see that the sub-component from the Cu-$4p$ states
(which actually originate mainly from a re-expansion of the O-$2p$ tails)
is more negative than the total $p$-$p$ and that the sub-component from the
low-lying ($\sim-5$ Ry) semicore Cu-$3p$ states is small and positive.
From this we could also determine that the more negative
PBE0 $p$-$p$ component come half and half from the Cu-$3p$ and Cu-$4p$ states.

Figure \ref{fig4} shows the density of states (DOS) of Cu$_{2}$O for a few
selected functionals. We can see that in the energy range between $-8$ and
$-5$ eV below the Fermi energy (set at $E=0$ eV), most of the DOS
comes from O-$2p$ electrons.
The DOS between $-4$ and 0 eV is entirely due to Cu-$3d$ states, while
above the band gap, the different partial DOSs actually represent
Cu-$4s$ states. By comparing the different functionals,
we can observe that the O-$2p$ peaks are higher in energy (closer to the
Cu-$3d$ states) for the FLL version of LDA+$U$.
Also, the LDA+$U$ and hybrid methods shift the main Cu-$3d$ peaks down
in energy.

Compared to the other hybrid results from the literature (also shown in
Table \ref{table4}), we can see that the HSE (with $\alpha_{\text{x}}=0.275$)
band gap of 2.12 eV\cite{ScanlonJPCL10}
is close to the YS-PBE0 value of 1.99 eV, as expected from the results obtained in
Sec. \ref{YSPBE}. The B3LYP band gap of 2.1 eV reported in
Ref. \onlinecite{HuPRB08} is much smaller than our PBE0 value of 2.77 eV.
This is mainly due to the smaller amount of HF exchange $\alpha_{\text{x}}$ in
B3LYP (0.2 for B3LYP versus 0.25 for PBE0).

\section{\label{summary}Summary}

The implementation of unscreened and screened hybrid functionals into
the \textsc{wien2k} code, which is based on the LAPW basis set,
has been presented. The screening is based on the Yukawa potential for
which the expansion in spherical harmonics has a simple expression.
Also, it was possible to calculate analytically all integrals which were
necessary for the derivation of the various formulas for the pseudocharge
method. In order to check the validity of the implementation, first, test
calculations were done on systems which do not contain core electrons, such
that the total Hartree-Fock energy could be compared with benchmark results
from the literature. As a further test of the implementation, the band
gap and lattice constant of several solids have been calculated with the
hybrid functionals PBE0 and YS-PBE0 which are based on the GGA functional PBE.
The results are in very good agreement
with the results obtained by other codes. Noticeably, for the screened hybrid
functional YS-PBE0, it was possible to find a value of the screening parameter
$\lambda$ such that the results are very close to the results of HSE06, whose
screening is based on the error function. Finally, we applied the hybrid
functionals to the semiconductor Cu$_{2}$O. The results obtained
with the unscreened PBE0 and screened YS-PBE0 for the band gap and EFG are
much more accurate than the results obtained with semilocal and LDA+$U$
functionals.

\appendix

\section{\label{Pseudocharge}Pseudocharge method}

In this appendix, the basic formulas of the pseudocharge method
are given (Sec. \ref{basic}), as well as explicit expressions
used for the HF energy (Sec. \ref{formulasHF}).

\subsection{\label{basic}Basic formulas}

In all-electron calculations, the charge density $\rho$ has large
oscillations near the nuclei, therefore
its Fourier expansion will converge slowly, making the calculation of the
potential $v$ generated by $\rho$ with Fourier transforms inefficient.
The idea of the pseudocharge method\cite{WeinertJMP81} is to
replace the charge density $\rho$ inside the atomic spheres $\text{S}_{\alpha}$
by a smoother one ($\tilde{\rho}$) such that the Fourier expansion of
$\tilde{\rho}=\rho_{\text{PW}}+\overline{\rho}$ converges
faster. $\rho_{\text{PW}}$ is the continuation inside the spheres
of the plane waves (PW) representation of the charge density
and $\overline{\rho}=\sum_{\alpha}\overline{\rho}_{\alpha}$
is zero in the interstitial region. Such a scheme is possible since the
potential in the interstitial region created by the charge inside the spheres
depends only on the multipole moments $q_{\ell m}^{\alpha}$ which are defined
as follows for the unscreened [Eq. (\ref{vc})] and screened [Eq. (\ref{vcs})]
potentials:
\begin{equation}
q_{\ell m}^{\alpha} =
\int\limits_{\text{S}_{\alpha}}
Y_{\ell m}^{*}\left(\hat{\mathbf{r}}\right)r^{\ell}
\rho\left(\mathbf{r}\right)d^{3}r,
\label{qlm}
\end{equation}
\begin{equation}
q_{\ell m}^{\alpha} =
\frac{\left(2\ell+1\right)!!}{\lambda^{\ell}}\int\limits_{\text{S}_{\alpha}}
Y_{\ell m}^{*}\left(\hat{\mathbf{r}}\right)i_{\ell}\left(\lambda r\right)
\rho\left(\mathbf{r}\right)d^{3}r.
\label{qlms}
\end{equation}
Therefore, $\overline{\rho}_{\alpha}$ should be chosen such that
inside the spheres, the multipole moments of $\tilde{\rho}$ are equal
to the multipole moments of the true charge density $\rho$.
After having determined $\tilde{\rho}$, the potential in the interstital region
is calculated with
\begin{equation}
v_{\text{I}}(\mathbf{r}) =
4\pi\sum_{\mathbf{G}}\frac{\tilde{\rho}_{\mathbf{G}}}
{\left\vert\mathbf{G}\right\vert^{2}}e^{\text{i}\mathbf{G}\cdot\mathbf{r}}
\label{vI}
\end{equation}
and
\begin{equation}
v_{\text{I}}(\mathbf{r}) =
4\pi\sum_{\mathbf{G}}\frac{\tilde{\rho}_{\mathbf{G}}}
{\left\vert\mathbf{G}\right\vert^{2}+\lambda^{2}}
e^{\text{i}\mathbf{G}\cdot\mathbf{r}}
\label{vIs}
\end{equation}
for the unscreened and screened potentials,
respectively. Then, inside the atomic sphere S$_{\alpha}$, the potential is
the solution of a Green function problem:
\begin{eqnarray}
v_{\alpha}(\mathbf{r}) & = &
\int\limits_{\text{S}_{\alpha}}\rho(\mathbf{r}')
G^{\alpha}\left(\mathbf{r},\mathbf{r}'\right)d^{3}r' \nonumber \\
& & -\frac{R_{\alpha}^{2}}{4\pi}\oint\limits_{\text{S}_{\alpha}}
v_{\text{I}}\left(\mathbf{r}'\right)\frac{\partial G^{\alpha}}{\partial n'}
\left(\mathbf{r},\mathbf{r}'\right)\sin\theta'd\theta'd\phi',
\label{va}
\end{eqnarray}
where the Green function is given by
\cite{WeinertJMP81,Jackson}
\begin{equation}
G^{\alpha}\left(\mathbf{r},\mathbf{r}'\right) =
\sum_{\ell=0}^{\infty}\sum_{m=-\ell}^{\ell}
G_{\ell}^{\alpha}\left(r,r'\right)
Y_{\ell m}^{*}(\hat{\mathbf{r}}')Y_{\ell m}(\hat{\mathbf{r}}),
\label{Ga}
\end{equation}
where
\begin{equation}
G_{\ell}^{\alpha}(r,r') =
\frac{4\pi}{2\ell+1}\frac{r_{<}^{\ell}}{r_{>}^{\ell+1}}
\left(1 - \frac{r_{>}^{2\ell+1}}{R_{\alpha}^{2\ell+1}}\right)
\label{Gla}
\end{equation}
or
\begin{eqnarray}
G_{\ell}^{\alpha}(r,r') & = &
4\pi\lambda
i_{\ell}\left(\lambda r_{<}\right)k_{\ell}\left(\lambda r_{>}\right) \nonumber \\
& & \times
\left(1 -
\frac{k_{\ell}\left(\lambda R_{\alpha}\right)i_{\ell}\left(\lambda r_{>}\right)}
{i_{\ell}\left(\lambda R_{\alpha}\right)k_{\ell}\left(\lambda r_{>}\right)}
\right)
\label{Glas}
\end{eqnarray}
for the unscreened and screened potentials, respectively.
$\partial G^{\alpha}/\partial n'$ is the normal
derivative of $G^{\alpha}$ at the sphere boundary.

\subsection{\label{formulasHF}Explicit expressions for the Hartree-Fock energy}

In Eqs. (\ref{vnkG}) and (\ref{vnkGs}), the Fourier coefficients of the pseudocharge
density are given by
\begin{equation}
\tilde{\rho}_{n\mathbf{k}n'\mathbf{k}'}^{\sigma\mathbf{q}} =
\rho_{n\mathbf{k}n'\mathbf{k}'}^{\sigma\mathbf{G}} +
\overline{\rho}_{n\mathbf{k}n'\mathbf{k}'}^{\sigma\mathbf{q}},
\label{qqq}
\end{equation}
where (unscreened case)
\begin{eqnarray}
\overline{\rho}_{n\mathbf{k}n'\mathbf{k}'}^{\sigma\mathbf{q}} & = &
\frac{4\pi}{\Omega}\sum_{\alpha}^{\text{cell}}\sum_{\ell,m}
\frac{(2\ell+2p+3)!!}{R_{\alpha}^{\ell+p+1}}
\frac{(-\text{i})^{\ell}}{(2\ell+1)!!} \nonumber \\
& & \times\frac{j_{\ell+p+1}\left(\left\vert\mathbf{q}\right\vert
R_{\alpha}\right)}{\left\vert\mathbf{q}\right\vert^{p+1}}
e^{-\text{i}\mathbf{q}\cdot\tau_{\alpha}}
Y_{\ell m}\left(\widehat{\mathbf{q}}\right)
\overline{q}_{\ell m}^{\alpha\sigma n\mathbf{k}n'\mathbf{k}'}\nonumber\\
\label{rhopseudo}
\end{eqnarray}
or (screened case)
\begin{eqnarray}
\overline{\rho}_{n\mathbf{k}n'\mathbf{k}'}^{\sigma\mathbf{q}} & = &
\frac{4\pi}{\Omega}\sum_{\alpha}^{\text{cell}}\sum_{\ell,m}
\frac{\lambda^{\ell+p+1}}{i_{\ell+p+1}\left(\lambda R_{\alpha}\right)}
\frac{(-\text{i})^{\ell}}{(2\ell+1)!!} \nonumber \\
& & \times\frac{j_{\ell+p+1}\left(\left\vert\mathbf{q}\right\vert
R_{\alpha}\right)}{\left\vert\mathbf{q}\right\vert^{p+1}}
e^{-\text{i}\mathbf{q}\cdot\tau_{\alpha}}
Y_{\ell m}\left(\widehat{\mathbf{q}}\right)
\overline{q}_{\ell m}^{\alpha\sigma n\mathbf{k}n'\mathbf{k}'}.\nonumber\\
\label{rhopseudos}
\end{eqnarray}
In Eqs. (\ref{rhopseudo}) and (\ref{rhopseudos}), $p$ is an
integer which is chosen such that $\ell+p$ is fixed.\cite{WeinertJMP81}
For $\mathbf{q}=\mathbf{0}$,
Eqs. (\ref{rhopseudo}) and (\ref{rhopseudos}) reduce to
\begin{equation}
\overline{\rho}_{n\mathbf{k}n'\mathbf{k}}^{\sigma\mathbf{0}} =
\frac{\sqrt{4\pi}}{\Omega}\sum_{\alpha}^{\text{cell}}
\overline{q}_{00}^{\alpha\sigma n\mathbf{k}n'\mathbf{k}}
\label{rho0}
\end{equation}
and
\begin{equation}
\overline{\rho}_{n\mathbf{k}n'\mathbf{k}}^{\sigma\mathbf{0}} =
\frac{\sqrt{4\pi}}{\Omega}\sum_{\alpha}^{\text{cell}}
\frac{\left(\lambda R_{\alpha}\right)^{p+1}}{\left(2p+3\right)!!
i_{p+1}\left(\lambda R_{\alpha}\right)}
\overline{q}_{00}^{\alpha\sigma n\mathbf{k}n'\mathbf{k}},
\label{rho0s}
\end{equation}
respectively.
In Eqs. (\ref{rhopseudo})-(\ref{rho0s}),
\begin{equation}
\overline{q}_{\ell m}^{\alpha\sigma n\mathbf{k}n'\mathbf{k}'} =
q_{\ell m}^{\alpha\sigma n\mathbf{k}n'\mathbf{k}'} -
q_{\ell m}^{\text{PW},\alpha\sigma n\mathbf{k}n'\mathbf{k}'},
\label{qlmbar}
\end{equation}
where $q_{\ell m}^{\alpha\sigma n\mathbf{k}n'\mathbf{k}'}$ and
$q_{\ell m}^{\text{PW},\alpha\sigma n\mathbf{k}n'\mathbf{k}'}$ are
the multipole moments of
$\rho_{n\mathbf{k}n'\mathbf{k}'}^{\sigma}$ inside the spheres and of
the continuation of the PW representation of
$\rho_{n\mathbf{k}n'\mathbf{k}'}^{\sigma}$ inside the spheres,
respectively, whose expressions are given by
\begin{eqnarray}
q_{\ell m}^{\alpha\sigma n\mathbf{k}n'\mathbf{k}'} & = &
\sum_{\ell_{1},\ell_{2}}\sum_{f_{1},f_{2}}
T_{\alpha\sigma n\mathbf{k}n'\mathbf{k}'}^{f_{1}f_{2}\ell_{1}\ell_{2}\ell m} \nonumber \\
& & \times\int\limits_{0}^{R_{\alpha}}
u_{f_{1}\ell_{1}}^{\alpha\sigma}(r)u_{f_{2}\ell_{2}}^{\alpha\sigma}(r)
r^{\ell+2}dr
\label{qlmnk}
\end{eqnarray}
\begin{eqnarray}
q_{\ell m}^{\text{PW},\alpha\sigma n\mathbf{k}n'\mathbf{k}'} & = &
\sum_{\mathbf{G}}
\frac{4\pi\text{i}^{\ell}R_{\alpha}^{\ell+2}
j_{\ell+1}\left(\left\vert\mathbf{q}\right\vert R_{\alpha}\right)}
{\left\vert\mathbf{q}\right\vert} \nonumber \\
& & \times e^{\text{i}\mathbf{q}\cdot\tau_{\alpha}}
Y_{\ell m}^{*}\left(\widehat{\mathbf{q}}\right)
\rho_{n\mathbf{k}n'\mathbf{k}'}^{\sigma\mathbf{G}}
\label{qlmnkPW}
\end{eqnarray}
for the unscreened potential and
\begin{eqnarray}
q_{\ell m}^{\alpha\sigma n\mathbf{k}n'\mathbf{k}'} & = &
\frac{\left(2\ell+1\right)!!}{\lambda^{\ell}}
\sum_{\ell_{1},\ell_{2}}\sum_{f_{1},f_{2}}
T_{\alpha\sigma n\mathbf{k}n'\mathbf{k}'}^{f_{1}f_{2}\ell_{1}\ell_{2}\ell m} \nonumber \\
& & \times\int\limits_{0}^{R_{\alpha}}
u_{f_{1}\ell_{1}}^{\alpha\sigma}(r)u_{f_{2}\ell_{2}}^{\alpha\sigma}(r)
i_{\ell}\left(\lambda r\right)r^{2}dr
\label{qlmnks}
\end{eqnarray}
\begin{widetext}
\begin{equation}
q_{\ell m}^{\text{PW},\alpha\sigma n\mathbf{k}n'\mathbf{k}'} =
\sum_{\mathbf{G}}
\left[\lambda j_{\ell}\left(\left\vert\mathbf{q}\right\vert
R_{\alpha}\right)i_{\ell-1}\left(\lambda R_{\alpha}\right) -
\left\vert\mathbf{q}\right\vert
j_{\ell-1}\left(\left\vert\mathbf{q}\right\vert
R_{\alpha}\right)i_{\ell}\left(\lambda R_{\alpha}\right)\right]
\frac{4\pi\text{i}^{\ell}R_{\alpha}^{2}\left(2\ell+1\right)!!} 
{\lambda^{\ell}\left(\left\vert\mathbf{q}\right\vert^{2}+\lambda^{2}\right)}
e^{\text{i}\mathbf{q}\cdot\tau_{\alpha}}
Y_{\ell m}^{*}\left(\widehat{\mathbf{q}}\right)
\rho_{n\mathbf{k}n'\mathbf{k}'}^{\sigma\mathbf{G}}
\label{qlmnkPWs}
\end{equation}
\end{widetext}
for the screened potential. When $\mathbf{k}=\mathbf{k}'$, the term
$\mathbf{G}=\mathbf{0}$ in Eqs. (\ref{qlmnkPW}) and (\ref{qlmnkPWs}) reduces to
\begin{equation}
\delta_{\ell0}\sqrt{4\pi}
\frac{R_{\alpha}^{3}}{3}
\rho_{n\mathbf{k}n'\mathbf{k}}^{\sigma\mathbf{0}}
\label{qlmnk0}
\end{equation}
and
\begin{equation}
\delta_{\ell0}\sqrt{4\pi}
\frac{R_{\alpha}^{2}i_{1}\left(\lambda R_{\alpha}\right)}{\lambda}
\rho_{n\mathbf{k}n'\mathbf{k}}^{\sigma\mathbf{0}},
\label{qlmnks0}
\end{equation}
respectively.

\section{\label{totalenergy}Total energy}

For the case of a hybrid exchange-correlation functional,
the total energy is given by (spin-unpolarized form)
\begin{eqnarray}
E_{\text{tot}} & = &
T_{\text{s}} +
\frac{1}{2}\int\limits_{\text{cell}}
v_{\text{Coul}}(\mathbf{r})\rho(\mathbf{r})d^{3}r -
\frac{1}{2}\sum_{\alpha}^{\text{cell}}Z_{\alpha}
v_{\text{M}}^{\alpha}(\tau_{\alpha}) \nonumber \\
& & +
E_{\text{xc}}^{\text{SL}} + \alpha_{\text{x}}\left(
E_{\text{x}}^{\text{HF}} - E_{\text{x}}^{\text{SL}}\right), \nonumber \\
\label{Etot1}
\end{eqnarray}
where $T_{\text{s}}$ is the kinetic energy of the electrons and
\begin{equation}
v_{\text{Coul}}(\mathbf{r}) =
\int\limits_{\text{crystal}}\frac{\rho(\mathbf{r}')}
{\left\vert\mathbf{r}-\mathbf{r}'\right\vert}d^{3}r' -
\sum_{\beta}^{\text{crystal}}
\frac{Z_{\beta}}{\left\vert\mathbf{r}-\tau_{\beta}\right\vert},
\label{vcoul}
\end{equation}
\begin{equation}
v_{\text{M}}^{\alpha}(\tau_{\alpha}) =
\int\limits_{\text{crystal}}\frac{\rho(\mathbf{r}')}
{\left\vert\tau_{\alpha}-\mathbf{r}'\right\vert}d^{3}r' -
\sum_{\beta\atop\beta\neq\alpha}^{\text{crystal}}
\frac{Z_{\beta}}{\left\vert\tau_{\alpha}-\tau_{\beta}\right\vert},
\label{vM}
\end{equation}
are the Coulomb and Madelung potentials, respectively. By using the sum of the
eigenvalues
\begin{widetext}
\begin{equation}
\sum_{n_{\text{c}},\ell_{\text{c}},m_{\text{c}}}
\epsilon_{n_{\text{c}}\ell_{\text{c}}m_{\text{c}}} +
\sum_{n,\mathbf{k}}w_{n\mathbf{k}}\epsilon_{n\mathbf{k}} =
T_{\text{s}} +
\int\limits_{\text{cell}}
v_{\text{Coul}}(\mathbf{r})\rho(\mathbf{r})d^{3}r +
\int\limits_{\text{cell}}v_{\text{xc}}^{\text{SL}}(\mathbf{r})
\rho(\mathbf{r})d^{3}r + 
\alpha_{\text{x}}\left(
2E_{\text{x,vv}}^{\text{HF}} + E_{\text{x,vc}}^{\text{HF}} -
\int\limits_{\text{cell}}
v_{\text{x}}^{\text{SL}}(\mathbf{r})\rho_{\text{val}}(\mathbf{r})d^{3}r
\right),
\label{sumNL1}
\end{equation}
\end{widetext}
where $\rho_{\text{val}}$ is the valence electron density
(the core electrons experience the semilocal potential),
the total energy can be rewritten as
\begin{widetext}
\begin{eqnarray}
E_{\text{tot}}
 & = &
\sum_{n_{\text{c}},\ell_{\text{c}},m_{\text{c}}}
\epsilon_{n_{\text{c}}\ell_{\text{c}}m_{\text{c}}} +
\sum_{n,\mathbf{k}}w_{n\mathbf{k}}\epsilon_{n\mathbf{k}}
-\frac{1}{2}\int\limits_{\text{cell}}
v_{\text{Coul}}(\mathbf{r})\rho(\mathbf{r})d^{3}r -
\frac{1}{2}\sum_{\alpha}^{\text{cell}}Z_{\alpha}
v_{\text{M}}^{\alpha}(\tau_{\alpha}) -
\int\limits_{\text{cell}}
v_{\text{xc}}^{\text{SL}}(\mathbf{r})\rho(\mathbf{r})d^{3}r
\nonumber\\
& & +
E_{\text{xc}}^{\text{SL}} + \alpha_{\text{x}}\left(
E_{\text{x,cc}}^{\text{HF}} - E_{\text{x,vv}}^{\text{HF}} +
\int\limits_{\text{cell}}
v_{\text{x}}^{\text{SL}}(\mathbf{r})\rho_{\text{val}}(\mathbf{r})d^{3}r -
E_{\text{x}}^{\text{SL}}\right).
\label{Etot2}
\end{eqnarray}
\end{widetext}
For a screened hybrid functional, the exchange-only terms are simply
replaced by their SR counterparts.
The use of the second variational procedure allows us to write
the sum of the valence eigenvalues in the following way:
\begin{eqnarray}
\sum_{n,\mathbf{k}}w_{n\mathbf{k}}\epsilon_{n\mathbf{k}} & = &
\sum_{n,\mathbf{k}}w_{n\mathbf{k}}
\sum_{m}\left\vert c_{n\mathbf{k}}^{m}\right\vert^{2}
\epsilon_{m\mathbf{k}}^{\text{SL}} + \alpha_{\text{x}}\left(
2E_{\text{x,vv}}^{\text{HF}} \right.\nonumber \\
& &
\left. +
E_{\text{x,vc}}^{\text{HF}} -
\int\limits_{\text{cell}}
v_{\text{x}}^{\text{SL}}(\mathbf{r})
\rho_{\text{val}}(\mathbf{r})d^{3}r
\right),
\label{sumNL2}
\end{eqnarray}
where $c_{n\mathbf{k}}^{m}$ are the coefficients of the expansion of
$\psi_{n\mathbf{k}}$ ($\psi_{n\mathbf{k}} =
\sum_{m}c_{n\mathbf{k}}^{m}\psi_{m\mathbf{k}}^{\text{SL}}$).
From Eq. (\ref{sumNL2}),
the valence-valence HF exchange energy $E_{\text{x,vv}}^{\text{HF}}$
can be calculated, thus avoiding the use of Eq. (\ref{ExvvHF}), which
is the most expensive component of the total energy to calculate.

\section{\label{Functionalderivative}Functional derivative of
$E_{\text{x}}^{\text{SR-SL}}$}

The functional derivative of $E_{\text{x}}^{\text{SR-SL}}$
[Eq. (\ref{ExSRSL})] for the spin-unpolarized case is given by
\begin{equation}
v_{\text{x}}^{\text{SR-SL}} =
-\frac{3}{4}\left(\frac{3}{\pi}\right)^{1/3}\left(v_{1}J + v_{2}\frac{dJ}{da} +
v_{3}\frac{d^{2}J}{da^{2}}\right),
\label{VXSRSL}
\end{equation}
where
\begin{equation}
v_{1} =
\frac{4}{3}\rho^{1/3}F_{\text{x}} -
\frac{1}{b^{2}}\frac{\nabla^{2}\rho}{\rho^{4/3}}H_{\text{x}} +
\left(
\frac{4}{3}\rho^{1/3}s^{3} -
\frac{1}{b^{3}}\frac{t}{\rho^{8/3}}
\right)
\frac{dH_{\text{x}}}{ds},
\label{v1}
\end{equation}
\begin{eqnarray}
v_{2} & = &
-\frac{1}{3}\frac{\lambda}{b}F_{\text{x}}^{3/2} +
\left(
\frac{1}{2}\frac{\lambda}{b}s^{2} -
\frac{1}{2}\frac{\lambda}{b^{3}}\frac{\nabla^{2}\rho}{\rho^{5/3}}
\right)
F_{\text{x}}^{1/2}H_{\text{x}} \nonumber \\
 & & +
\left(
\frac{2}{3}\frac{\lambda}{b}s^{3} -
\frac{1}{2}\frac{\lambda}{b^{4}}
\frac{t}{\rho^{3}}
\right)
F_{\text{x}}^{1/2}\frac{dH_{\text{x}}}{ds} \nonumber \\
& & +
\left(
\frac{\lambda}{b}{s^{4}} -
\frac{3}{4}\frac{\lambda}{b^{4}}
\frac{st}{\rho^{3}}
\right)
\frac{H_{\text{x}}^{2}}{F_{\text{x}}^{1/2}},
\label{v2}
\end{eqnarray}
\begin{equation}
v_{3} =
\frac{1}{6}\frac{\lambda^{2}}{b^{2}}\frac{s^{2}}{\rho^{1/3}}
F_{\text{x}}H_{\text{x}} +
\left(
\frac{1}{3}\frac{\lambda^{2}}{b^{2}}\frac{s^{4}}{\rho^{1/3}} -
\frac{1}{4}\frac{\lambda^{2}}{b^{5}}
\frac{st}{\rho^{10/3}}
\right)
H_{\text{x}}^{2}
\label{v3},
\end{equation}
where $b=2\left(3\pi^{2}\right)^{1/3}$,
$s=\left\vert\nabla\rho\right\vert/\left(2\left(3\pi^{2}\right)^{1/3}\rho^{4/3}\right)$,
$t=\nabla\rho\cdot\nabla\left\vert\nabla\rho\right\vert$, and
$H_{\text{x}}=\left(1/s\right)dF_{\text{x}}/ds$.

\begin{acknowledgments}

We are grateful to Robert Laskowski, Sandro Massidda, and Michael Weinert for
very useful discussions. This work was supported by the projects P20271-N17
and SFB-F41 (ViCoM) of the Austrian Science Fund.

\end{acknowledgments}

\end{document}